\documentclass[aps,prc,twocolumn,showpacs,showkeys,amsmath,amssymb,nofootinbib]{revtex4-1}

\usepackage{graphicx}
\usepackage{dcolumn}
\usepackage{bm}

\usepackage{siunitx}
\usepackage{isotope}
\usepackage{color}

\usepackage{afterpage}

\begin{document}

\title{Comparison of ultracold neutron sources for fundamental physics measurements}


\author{G.~Bison}
\affiliation{Laboratory for Particle Physics, Paul Scherrer Institute (PSI), Villigen, Switzerland}

\author{T.~Brenner}
\affiliation{Institut Laue Langevin (ILL), Grenoble, France}

\author{M.~Daum}
\affiliation{Laboratory for Particle Physics, Paul Scherrer Institute (PSI), Villigen, Switzerland}

\author{M.~Beck}
\affiliation{Institute of Physics, Johannes Gutenberg University, Mainz, Germany}

\author{K.~Eberhardt}
\affiliation{Institute of Nuclear Chemistry, Johannes Gutenberg University, Mainz, Germany}

\author{P.~Geltenbort}
\affiliation{Institut Laue Langevin (ILL), Grenoble, France}

\author{C.~Geppert}
\affiliation{Institute of Nuclear Chemistry, Johannes Gutenberg University, Mainz, Germany}

\author{W.~Heil}
\affiliation{Institute of Physics, Johannes Gutenberg University, Mainz, Germany}

\author{T.~Jenke}
\affiliation{Institut Laue Langevin (ILL), Grenoble, France}

\author{J.~Kahlenberg}
\affiliation{Institute of Physics, Johannes Gutenberg University, Mainz, Germany}

\author{J.~Karch}
\affiliation{Institute of Physics, Johannes Gutenberg University, Mainz, Germany}

\author{S.~Karpuk}
\affiliation{Institute of Nuclear Chemistry, Johannes Gutenberg University, Mainz, Germany}

\author{K.~Kirch}
\altaffiliation{also at: Institute for Particle Physics, Eidgen\"ossische Technische Hochschule (ETH), Z\"urich, Switzerland}
\affiliation{Laboratory for Particle Physics, Paul Scherrer Institute (PSI), Villigen, Switzerland}

\author{B.~Lauss}
\email[corresponding author:]{bernhard.lauss@psi.ch}
\affiliation{Laboratory for Particle Physics, Paul Scherrer Institute (PSI), Villigen, Switzerland}

\author{T.~Reich}
\affiliation{Institute of Nuclear Chemistry, Johannes Gutenberg University, Mainz, Germany}

\author{D.~Ries}
\email[corresponding author:]{dieter.ries@psi.ch}
\altaffiliation{also at: Institute for Particle Physics, Eidgen\"ossische Technische Hochschule (ETH), Z\"urich, Switzerland}
\affiliation{Laboratory for Particle Physics, Paul Scherrer Institute (PSI), Villigen, Switzerland}

\author{K.~Ross}
\affiliation{Institute of Physics, Johannes Gutenberg University, Mainz, Germany}

\author{P.~Schmidt-Wellenburg}
\affiliation{Laboratory for Particle Physics, Paul Scherrer Institute (PSI), Villigen, Switzerland}

\author{C.~Siemensen}
\affiliation{Institute of Nuclear Chemistry, Johannes Gutenberg University, Mainz, Germany}

\author{Y.~Sobolev}
\affiliation{Institute of Nuclear Chemistry, Johannes Gutenberg University, Mainz, Germany}

\author{N.~Trautmann}
\affiliation{Institute of Nuclear Chemistry, Johannes Gutenberg University, Mainz, Germany}

\author{G.~Zsigmond}
\affiliation{Laboratory for Particle Physics, Paul Scherrer Institute (PSI), Villigen, Switzerland}

\author{O.~Zimmer}
\affiliation{Institut Laue Langevin (ILL), Grenoble, France}

\date{\today}

\begin{abstract}
Ultracold neutrons (UCNs) are key for precision studies of 
fundamental parameters of the neutron and in searches for new
CP violating processes or exotic interactions
beyond the Standard Model of particle physics.
The most prominent example is the search for a permanent electric dipole moment of the neutron (nEDM).
We have performed an experimental comparison of 
the leading UCN sources currently operating. 
We have used a 'standard' UCN storage bottle 
with a volume of 32~liters, 
comparable in size to 
nEDM experiments, 
which allows us to 
compare the
UCN density available at a given beam port.
\end{abstract}



\maketitle

\section{Introduction}

Neutrons with very low kinetic energies
are reflected under all angles of incidence 
from suitable material surfaces. 
They can therefore be confined in material bottles 
and stored for times of up to several hundreds of seconds,
limited ultimately by their beta-decay lifetime of $\sim$880\,s.
This unique feature makes UCNs ideal to study 
fundamental properties of the neutron 
in precision experiments.
Appropriate bottle materials have
neutron optical potentials (``Fermi potentials'')
of up to about 350\,neV~\cite{Golub1991}.
Such energies correspond to 
neutron velocities below $\sim$8\,m/s,
or temperatures below $\sim$4\,mK.
Hence they are termed ultracold neutrons or UCNs. 

After the pioneering work of research groups in Dubna and Gatchina, Russia,~\cite{Lushchikov1969}
and in Munich, Germany,~\cite{Steyerl1969} 
many of the important UCN physics results
in the last decades were achieved 
using UCNs from the so-called 'Steyerl-turbine'~\cite{Steyerl1986}
at the PF2 facility of the 
Institut Laue Langevin (ILL), Grenoble, France.
This was for a long time the only source with 
sufficient UCNs available.
At the beginning experiments concentrated on 
measurements of the 
free neutron lifetime ($\tau_n$)~\cite{Serebrov2005tau,Pichlmaier2010,Arzumanov2012,Wie2011} and
on the search for a permanent electric dipole moment of the
neutron (nEDM)~\cite{ramsey1982,Baker2006,Baker2011,SerebrovEDM2015,Pendlebury2015}.
In addition to the fundamental particle aspect,
$\tau_n$
is also an important quantity in 
understanding the primordial nucleo-synthesis and
contributes presently one of the larger uncertainties
to its description~\cite{BBnucleosynthesis2015}.
The search for an nEDM,
which would violate CP, the combined 
symmetries of charge conjugation (C) and parity (P),
is considered to be one of the most
promising experiments in particle physics~\cite{Raidal2008,Engel2013}
to contribute to the 
understanding of the matter-antimatter asymmetry of the Universe.


%
The scope of UCN experiments has been extended
and includes measurements of neutron decay 
correlations~\cite{Liu2010,Plaster2012,UCNA2013},
searches for a finite electrical charge 
of the neutron~\cite{Borisov1988,Plonka2010,Siemensen2015},
%
investigations of gravitational effects 
(see e.g. Refs. in~\cite{Golub1991,Ignatovich1990} 
or~\cite{Afach2015PRD,Afach2015PRL}),
and tests of the weak equivalence principle
for the neutron~\cite{Frank2009,Kulin2015,Kulin2016,Kulin2016b}
or the theory of neutron diffraction~\cite{Bushuev2016,Kulin2016b}.
Measurements of gravitationally bound quantum states of neutrons
have been used for example for a high sensitivity search 
of deviations from Newton's gravity law
on the sub-millimeter scale~\cite{GRANIT2002,Jenke2011,Jenke2014}.

The emergent discussion for dark matter and dark energy
triggered further use of 
UCNs to search for exotic physics beyond the
Standard Model of particle physics,
like searches for mirror matter~\cite{Ban2007,Serebrov2008mirror,Altarev2009Mirror},
for Lorentz violation effects~\cite{Altarev2009LI},
for exotic interactions induced by axion-like 
particles~\cite{Serebrov2009Axion,Serebrov2010Axion,Zimmer2010a,Antoniadis2011,Afach2015Exotic},
dark matter~\cite{Serebrov2011DM,Serebrov2012DM},
or probing dark energy~\cite{Pignol2015,Lemmel2015,Schmiedmayer2015}.
The sensitivities of all such UCN experiments 
depend directly on the total UCN statistics available in the measurement, 
hence on high UCN densities in sizeable storage vessels.

All precision experiments in fundamental physics
with neutrons have in common
the need for high counting statistics.
Also the most recent and best nEDM limit~\cite{Pendlebury2015}
is limited in statistics,
and for efforts to improve this limit
more UCNs are imperative.
Therefore, worldwide efforts to increase UCN intensities 
started in the 1990'ies~\cite{Pokotilovski1995,Serebrov1997}
and new, so-called 'superthermal' UCN sources~\cite{Golub1975}
based on superfluid helium or solid deuterium 
have become operating in recent years~\cite{Kirch2010,LANL2000,Frei2007,Zimmer2011,Lauss2012}.

The determination of the UCN density
at a UCN source strongly depends on various factors.
There are different 
types of measurement like 
UCN storage experiments
and UCN flux measurements.
UCN densities can be determined
inside the UCN production volume of the source, or 
in a vessel
outside the source at a beamport accessible to experiments.
The surface materials and geometry of the storage apparatus
have an important influence via 
the 
Fermi potentials
of surfaces and total volume of storage.
UCN detectors show different detection efficiencies for UCNs,
depending on the used technology, for example 
\isotope[3]{He} proportional counters,
\isotope[10]{B} based counters, 
e.g.,~commercially available
``Cascade'' UCN detectors using GEM technology
\footnote{CD-T Technology, Hans-Bunte Strasse 8-10, 69123 Heidelberg, Germany},
scintillators \cite{Wei2016},
\isotope[6]{Li} doped scintillation counters~\cite{Goeltl2012,Afach2015USSA}
or activation measurements.
Also the treatment of data differs and influences the result
depending on assumed 
corrections for detection efficiencies
or UCN transmission, or if 
UCN counts observed in 
storage measurements are extrapolated to time equal zero.

Therefore, 
up to now a comparison between different sources 
was not directly possible.
In order to start changing this situation 
a simple, robust, and mobile UCN 
storage bottle was constructed and used to measure UCN densities at all
operating UCN sources worldwide~\cite{Bison2016}.
The experimental setup, including the UCN 
detector, was identical for all measurements
and the resulting data were analyzed in a consistent way.
This allows for
a quantitative comparison of the UCN densities for a given reasonably large 
UCN storage experiment. 
It does not, however, provide
an optimized measurement for the reachable UCN densities at any 
UCN source.
It is obvious that at some sources higher 
UCN densities could be measured, 
if the test storage volume would be smaller
or would be made from a different surface material.

The operating UCN sources under investigation were one 
out of four beamlines (the so-called “EDM” beamline) at 
PF2 (ILL)~\cite{Steyerl1986}, Grenoble, France,
%
%
at the TRIGA reactor of the Johannes Gutenberg University
Mainz, Germany~\cite{Frei2007,Lauer2013,Karch2014}, 
at the Paul Scherrer Institute, Villigen, 
Switzerland~\cite{Anghel2009,Lauss2011,Lauss2012,Lauss2014},
and
the new superfluid helium SUN-2 UCN source at
ILL.\footnote{No publication on the SUN-2 source in its current configuration 
is yet available. 
See Ref.~\cite{PSW2015sun}
for experiments using SUN-2 in an earlier stage of development, 
and 
Refs.~\cite{Zimmer2010,Piegsa2014,Leung2016,Courtois2011} 
for works performed with a first prototype and the predecessor source.}

%

In January 2015 comparison measurements at
the solid deuterium based UCN source 
of the Los Alamos National Laboratory (LANL),
Los Alamos, USA,
were performed~\cite{Ries2016}, 
but due to the atypical performance of the UCN source at that time
in comparison to the one given in Refs.~\cite{LANL2000,Saunders2004,LANL2013}
the LANL UCN collaboration did not wish to include the results in this paper. 
In the meantime the LANL UCN source has been decommissioned and 
is in the process of being upgraded~\cite{Pattie2016}.

The previously reported UCN source at RCNP Osaka, Japan~\cite{Masuda2012}
was not operational anymore in 2015 
and therefore measurements were not possible.

An international collaboration of the groups operating these UCN sources was formed.
This paper reports on the jointly performed measurements and their results,
and reflects the present status of the running UCN sources.
There are worldwide efforts to further increase UCN intensities 
at superthermal sources with upgrades planned 
or new UCN sources under construction. 


\section{A standard storage bottle}

\begin{figure}[htb]
\begin{center}
{\hspace*{-5.5mm}
\includegraphics[width=0.46\textwidth]{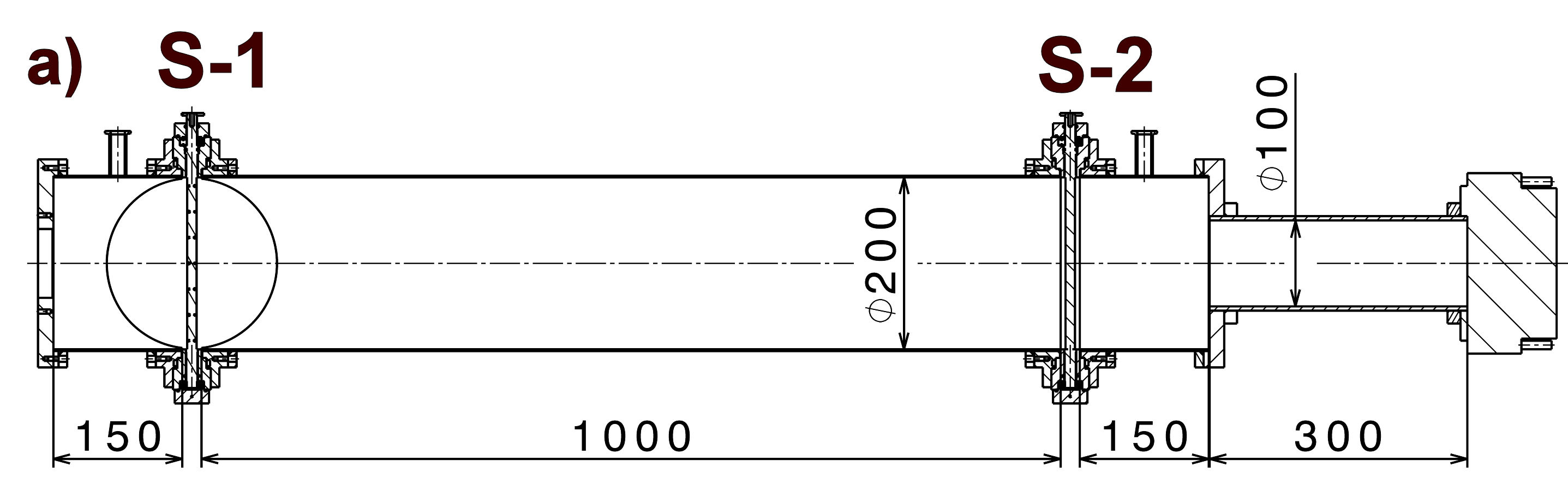}
}
\includegraphics[width=0.477\textwidth]{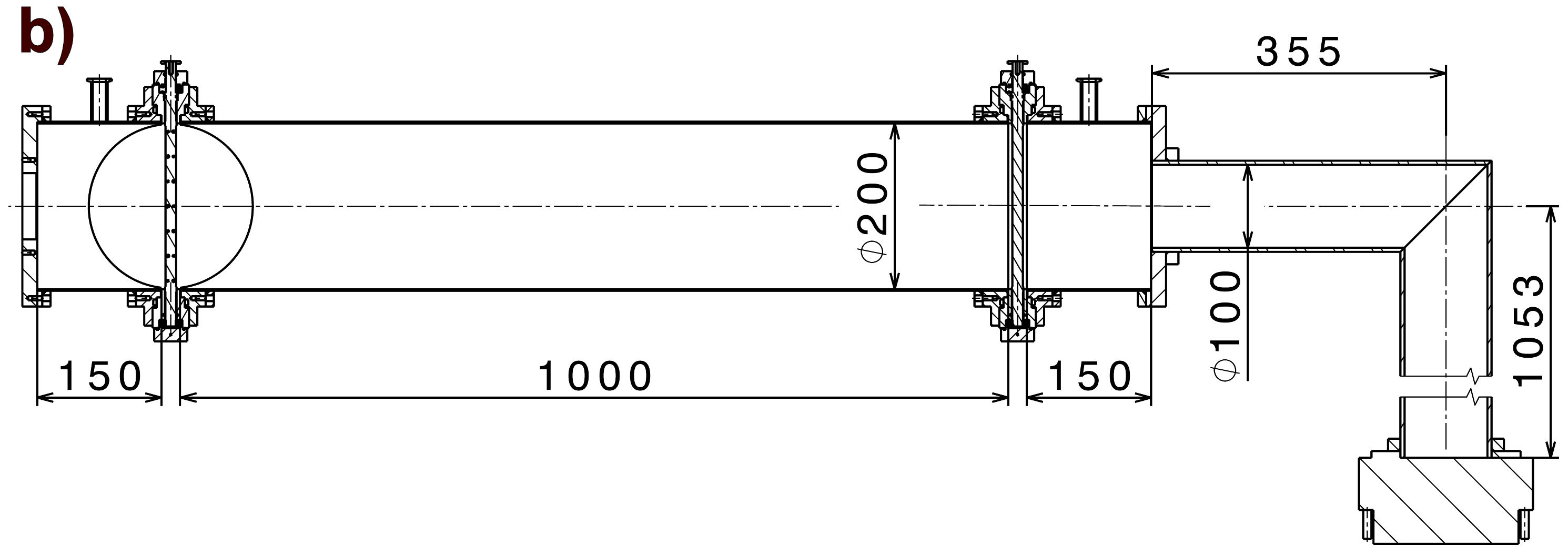}
\caption{
Side view of the UCN storage bottle with two different detector mounting options,
a) the horizontal extraction with straight UCN guide;
b) the vertical extraction with L-shaped UCN guide and the detector located 1\,m lower.
Visible from left to right are:
connection flange to source, shutter~1 (S-1), 
storage bottle,
shutter~2 (S-2), 
connection flange to detector, 
UCN guide, 
Cascade detector.
(All dimensions in mm).
}
\label{BottlePhoto}
\end{center}
\end{figure}

In order to measure the UCN density available at the beamport
of a given UCN source,
PSI developed a new storage setup
based on previous experiences of
determining UCN densities at the PSI UCN source~\cite{Bison2016}.
Fig.~\ref{BottlePhoto} shows drawings 
of the assembled storage bottle
with horizontal and vertical extraction
towards the detector.

This setup
consists of commercially available 
electro-polished stainless steel tubes
(complying with DIN-11866 standards)
and fast shutters without losses via openings 
during the filling and emptying movements.
The cylindrical UCN storage volume of the bottle 
has a length of 
\SI{1020.0\pm1.1}{\milli\meter},
and a resulting volume of \SI{32044\pm164}{\centi\meter^3},
including the error on the shutter dimension.
The overall storage volume size is
comparable to a typical storage vessel
e.g., employed in an nEDM measurement
and has a Fermi potential on the surface above 140\,neV.

The running nEDM experiment at PSI~\cite{Baker2011}
employs a deuterated polystyrene (dPS) coating in the
precession chamber with a Fermi potential of 161\,neV~\cite{Bodek2008}
and a volume of \SI{21000}{\centi\meter^3}.
The mobility of the setup
and hence a necessary certain robustness of construction
was an additional requirement.
The technical details and commissioning measurements
of the storage bottle are reported in Ref.~\cite{Bison2016}.

\section{The standardized measurement}


\subsection{The standard setup}

The storage experiment
was used in two variants for measurements 
at each source and location.
%
The adapter pieces,
the storage bottle itself
and the UCN detector stayed exactly the same for all measurements.
The guide used between the 
bottle and the detector was either a straight, \SI{300}{\milli\meter}
long acrylic tube coated with \isotope{NiMo} 85/15~\cite{Blau2016} 
or an similarly-coated L-shaped
acrylic tube with dimensions as shown in Fig.~\ref{BottlePhoto}.
The Fermi potential of the used NiMo coating was measured to be
\SI{220}{\nano\electronvolt}.
By connecting the detector
to the long arm and pointing it downwards
UCNs in the guide gain about 
\SI{110}{\nano\electronvolt} of kinetic energy
due to gravitational acceleration.
This increased energy allows the low-energy UCNs to pass through the detector's 
entrance window, which is made of
an aluminum foil of \SI{100}{\micro\meter} thickness and has
a reflective wall potential of \SI{54}{\nano\electronvolt}.

Comparing measurements done with the two different extraction 
guides towards the detector allows for a rough discrimination 
between the population of UCNs stored in the storage bottle with
kinetic energies below and above \SI{54}{\nano\electronvolt}.

In order to minimize
systematic errors, almost all peripheral hard- and software was identical in all
measurements, which meant that the very same modules were used.
This includes the UCN detector and all of its periphery,
i.e.,~readout electronics, trigger electronics, cables, high-voltage
power supply, data acquisition computer, and software.
The vacuum pumps 
and pressure gauges (Type Pfeiffer Vacuum PKR251)
used were shipped along with the setup.

After arrival and before the measurements at each source started,
the storage bottle was checked for gaps between individual components,
which could have opened up during transport. If such gaps were observed,
affected parts were reconnected with minimal gaps. During these 
inspections, the surfaces of the experiment were checked for dust or
other contaminants and cleaned if necessary, although this was barely 
necessary as the setup was transported in a closed state.
The setup was then brought into the same condition at every location 
which is determined by the geometry of the components.

Cascade detectors need a gas mixture of \isotope{Ar} and \isotope{CO_2}
with a volume ratio of \SI{70}{}-\SI{90}{\percent}\isotope{Ar} to
\SI{30}{}-\SI{10}{\percent}~\isotope{CO_2}.
At all UCN sources, gas from the local supplier was used. 
The specified
volume ratios were similar but not identical:
ILL: ``AIR LIQUIDE argon-CO$_2$ 90/10'', 
 \SI{90}{\percent}\isotope{Ar} to \SI{10}{\percent}\isotope{CO_2}.
TRIGA: ``Westfalen Gase Gasgemisch'', \SI{80}{\percent}\isotope{Ar} to \SI{20}{\percent}\isotope{CO_2}.
PSI: ``Messer Ferroline C18'', \SI{82}{\percent}\isotope{Ar} 
 to \SI{18}{\percent}\isotope{CO_2}.
The detector-gas flow through the detector was established at least \SI{24}{\hour}
before the first measurement started at all sources. 
The gas flow was manually
adjusted and checked in the same fashion.

During all measurements, the high voltage of the detector
was set to \SI{1350}{\volt}. 
The voltage was ramped directly 
to the target voltage without additional conditioning at higher voltage
being necessary.
The settings of the Cascade detector were identical in every measurement.
The time bin width was set to \SI{0.1}{\second}.
Specifically, the signal threshold voltages
{\texttt{Vref0}} and {\texttt{Vref1}}~\cite{Cascade2013}
were both
set to a value of 110\,a.~u. throughout all measurements.

\subsection{Standard measurement sequence}
\label{StandardSequence}

A single storage measurement consisted of the following sequence:

{\bf Start:} An electronic trigger signal, approximately coincident
with the start of UCN delivery, starts the timing sequence in the storage
experiment electronics.

{\bf Filling:} The source-side storage bottle shutter (shutter~1) is in
the open position to let UCNs coming from the source into the bottle. The 
detector-side storage bottle shutter (shutter~2) stays in the closed position.
This state lasts for the defined filling time.

{\bf Storage:} When the filling time is over, shutter~1
gets closes, and both shutters stay closed for the defined
storage time.

{\bf Counting:} 
During this period, shutter~1 stays closed, while shutter~2
is opened in order to empty the stored UCNs into the detector. 
The counting time was set to 100\,s at all UCN sources 
with the exception of
PF2, where 80\,s were used.

{\bf Trigger Ready:} After the counting time is over, the electronics is
reset to the initial state and then 
waits for the next trigger signal.\\

The standard procedures were the following for each 
extraction variant of the storage setup:

Optimization of the filling time
for the given source and measurement position:
This is done by performing storage measurements
where the filling time is being varied, but the storage and counting times stay
constant. 
The optimal filling time was defined as the shortest time with the highest output
of UCNs after \SI{5}{\second} of storage.

Measurements of the stored UCNs after storage times
of \SI{2}{}, \SI{5}{}, \SI{10}{}, \SI{20}{},
\SI{50}{}, 
\SI{100}{}, 
and \SI{200}{\second}
were performed, 
using the optimal filling time. 
The order in which the storage times were measured was 
randomized.
The measurements with \SI{200}{\second} storage time were typically performed 
only once as a consistency check.
The double exponential fit for determining the storage time constant (STC) 
only used data before 100\,s for statistics reasons.

The measurement of background and leakage rates 
is different for
any given UCN source, as the time structure of UCN delivery as well as the 
possibilities to close off the source vary widely.

\subsection{Standard analysis steps}
\label{StandardAnalysis}

The analysis of the measured UCN storage data is twofold.

 {\bf 1) Maximum UCN density:} 
The main objective of this measurement is 
the \emph{maximum UCN density} achievable under similar conditions at any UCN source.
The number of measured UCNs after a storage time of \SI{2}{\second} 
divided by the volume of 
the storage bottle is considered to be the maximum UCN density. 
No extrapolations
or scaling for inefficiencies in the measurement process are being applied.
Leaking UCNs from the source into the storage bottle during 
the storage and counting times have to be subtracted from the 
raw counted values because the shutters are not perfectly UCN tight. 
The leakage was measured for a \SI{2}{\second} storage time
measurement but with shutter~1 closed during the filling time period. 
The leakage counts for the other storage times 
were then calculated via an
exponential decrease using the measured longer storage time constant
(fit parameter $\tau_2$).
Background in the detector was measured at all sources and was found
to be below \SI{1}{\hertz} and therefore negligible. 
At all reactor
based UCN sources, cadmium shielding around the detector was used.

{\bf 2) Storage time constant:} 
The second observable of interest is the 
UCN lifetime inside the storage bottle, or the \emph{storage time constant}. 
This is primarily depending on the 
surface materials and quality and on the construction of the storage bottle,
namely gaps and cavities between individual pieces, and leakage of the shutter.
It also depends on the kinetic energies of the stored UCNs. 
 
As the storage bottle is exactly the same for all measurements, only the 
UCN energy dependence leads to variations of the storage time constant.
The lower the average energy, the larger
the storage time constant.
Hence, it can give a qualitative estimate of the average kinetic energies of UCNs
delivered by the respective source. 
 
A method to determine the storage time constant consists in determining 
the leakage of UCNs through shutter~2.
The rate of leakage is proportional to the amount of UCNs inside the bottle. 
The time dependence of this leakage rate for long storage times
reflects the time dependence of the UCN density inside the bottle and therefore 
provides an additional determination of
the storage time constant.
A double exponential model function is fitted to the rate of UCN leakage
\begin{equation}
N(t)\propto R(t) = A_1\times e^{-t/\tau_1} + A_2\times e^{-t/\tau_2}
\label{standardEquation}
\end{equation}
where $N(t)$ is the total amount of UCNs inside the storage bottle, $R(t)$ the
rate of leaked UCNs as measured in the detector, $A_1$ and $A_2$ are population
constants and $\tau_1$ and $\tau_2$ exponential decay time constants,
following the description of UCN losses as used in Ref.~\cite{Golub1991}.
The parameters $A_1$, $A_2$, $\tau_1$, and $\tau_2$ are determined by the fit.
$A_1$ and $\tau_1$ are considered to correspond to the population and lifetime
of UCNs with kinetic energies too high to be stored properly in the stainless-steel bottle. 
$\tau_1$ is typically below \SI{10}{\second}.
$A_2$ and $\tau_2$ correspond to storable UCNs, therefore $\tau_2$ is for
this analysis treated as the storage time constant to be compared between
UCN sources.

It is worth noting that the scope of this work is a comparison
of UCN sources, and neither a precision measurement of the maximally achievable
UCN density in an optimized storage volume for each given source nor a precision 
determination of the UCN lifetime in general or in the used storage bottle.

\section{Measurements at the superfluid He source SUN-2 at the Institut Laue Langevin}

The SUN-2 UCN source at the Institut Laue Langevin (ILL), Grenoble, France, 
is based on UCN production in superfluid $^4$He cooled to temperatures below 1\,K, 
where neutrons can become ultracold due to an inelastic single-phonon scattering process~\cite{Golub1977}. 
It converts cold neutrons with \SI{0.89}{\nano\meter} wavelength of a beam deflected 
by an intercalated-graphite monochromator~\cite{Courtois2011} to the beamline H172b with a 
flux density of \SI{9e8}{\centi\meter^{-2}\second^{-1}\nano\meter^{-1}}
at ``level~C'' of the ILL high-flux reactor.

The operating principle of the He-based source is detailed in~\cite{Leung2013,Leung2016} 
and in publications on the predecessor installation SUN-1, which employs similar 
concepts~\cite{Zimmer2011,Piegsa2014,PSW2015sun,Zimmer2007}. 
There, UCN densities of at least \SI{55}{UCNs\per\centi\meter^3} 
in the accumulation vessel were reported.

In the set-up used in this study, the cold neutron beam enters a converter vessel filled 
with superfluid helium with a volume of about 4\,litres. The volume of intersection 
with the beam is \SI{580}{} by \SI{80}{} by \SI{80}{\mm^3}.
The walls of the vessel are made from beryllium for entrance and exit of the beam, 
and beryllium-coated aluminium for the larger side walls. 
All internal surfaces are 
in addition coated with Fomblin grease (Solvay Solexis RT15). 
Beryllium has a Fermi potential of $V_F$=\SI{250}{\nano\electronvolt}, 
while the Fomblin grease has only about \SI{110}{\nano\electronvolt}. 
The cross section of UCNs for nuclear absorption in $^4$He is zero
and up-scattering 
in superfluid helium below 0.7~K
is negligibly small. 
UCNs can therefore be accumulated for very long times, 
with an ultimate limit set by the free neutron lifetime.

UCNs can be extracted from the superfluid helium through
a guide system detailed in Ref.~\cite{Leung2016}. 
A cold UCN valve situated at the exit of the converter vessel serves for accumulation and release of the UCNs. 
The opening in the helium container has a diameter of \SI{23}{\milli\meter}, 
followed by a conical guide section of \SI{100}{\milli\meter} length 
to a guide diameter of \SI{50}{\milli\meter}. 
\SI{280}{\milli\meter} above the gate valve
(\SI{380}{\milli\meter} above the lowest point in the UCN 
production vessel), the UCNs are fed into a horizontal stainless-steel UCN guide 
with an inner diameter of \SI{50}{\milli\meter}. 
If the Be sublayer under the Fomblin were not effective at all 
(or if the UCNs with energy beyond the Fomblin cutoff were completely 
removed from the vessel due to multiple passages through the bulk of the Fomblin), 
the UCN spectrum would thus be limited to about \SI{72}{\nano\electronvolt}. 
Close to the exit of the source another conical section tapers 
the guide to \SI{66}{\milli\meter} inner diameter.

\subsection{Setup at SUN-2}

\begin{figure}[htb]
\begin{center}
\includegraphics[width=0.5\textwidth]{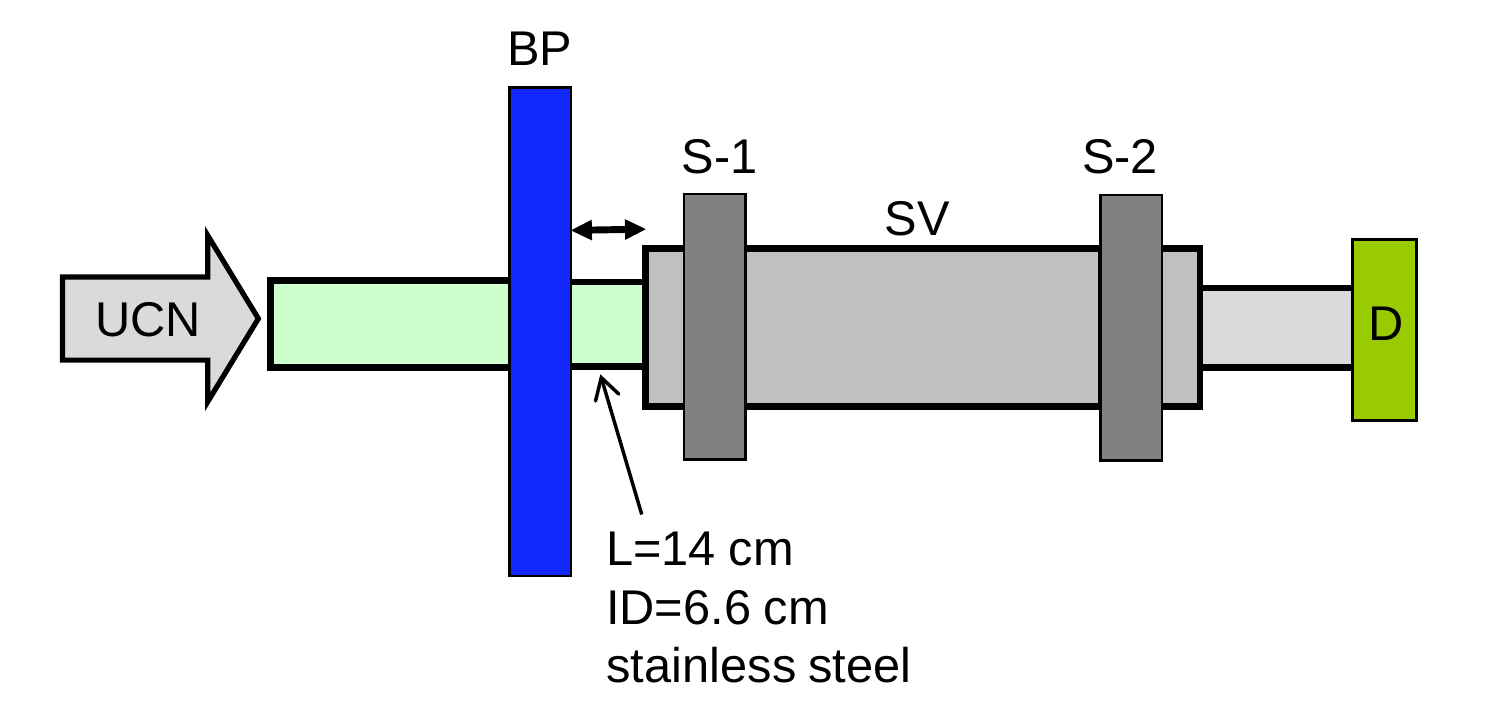}
\end{center}
\caption{Sketch 
of the setup at SUN-2 (not to scale). 
Indicated are the position of the beamport (BP), the length, 
diameter and materials of the connecting guides, and the direction of the UCNs.
The storage vessel is sketched with shutters S-1 and S-2, 
storage volume SV and detector D, as detailed in Fig.~\ref{BottlePhoto}(a) 
for the horizontal UCN extraction setup.
}
\label{SUN2setup}
\end{figure}

\begin{figure}[htb]
\begin{center}
\includegraphics[width=0.5\textwidth]{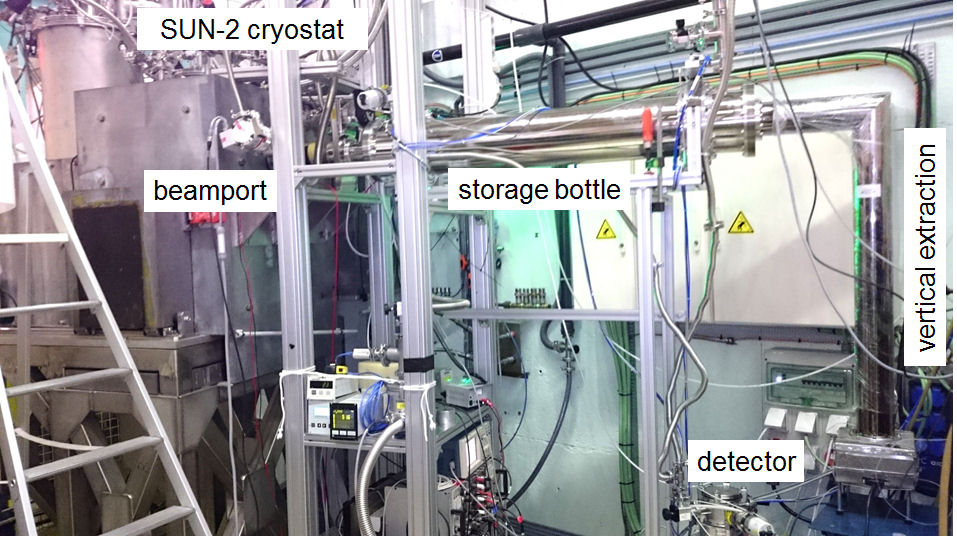}
\end{center}
\caption{Setup installed at the SUN-2 source with vertical extraction towards the detector. 
The source cryostat is visible on the left.}
\label{SUN2photos}
\end{figure}

The storage bottle was mounted directly at the output of the UCN source, 
using a \SI{141.5}{\milli\meter} long stainless-steel 
UCN guide with an inner diameter of \SI{66}{\milli\meter} (see Fig.~\ref{SUN2setup}). 
Fig.~\ref{SUN2photos} shows the storage bottle connected to the SUN-2 source, 
with the UCN detector mounted in the vertical extraction 1\,m lower. 
Measurements where also performed using the horizontal extraction guide.

While the vacuum shutter to the source was closed, the rest gas pressure in the storage bottle 
was typically in the high \SI{e-5}{\milli\bar} range but never exceeding \SI{2e-4}{\milli\bar}. 
As soon as the source shutter was opened the vacuum gauge readings were disturbed by the helium 
coming from the source due to the absence of a vacuum separation window between the storage bottle 
and the superfluid helium in the converter vessel. 
As a function of the state of the UCN valve, 
pressures fluctuating in the range between \SI{e-4}{\milli\bar} 
and \SI{e-2}{\milli\bar} were then observed.

\subsection{Operating conditions during the measurements}

The measurements were performed during the reactor 
cycle no. 175 at \SI{52.5\pm0.3}{\mega\watt} reactor power
in July 2015. 
The SUN-2 source was operated at helium temperatures 
in the range 0.65-\SI{0.7}{\kelvin}
(note that corrections to count rates due to the variations 
in temperature are negligible in this range).

Experiments were performed with UCN accumulation times of \SI{300}{\second} and \SI{600}{\second}, 
defined as the periods of time between the start of two consecutive UCN extractions 
to the external storage vessel (the UCN valve was kept closed in between extractions). 
Note that, in contrast to the definition of ``buildup mode'' measurements 
in earlier experiments~\cite{Zimmer2007,Zimmer2010}, 
the neutron beam stayed continuously on the converter vessel. 
The beam was switched off only for longer breaks in the measurement, e.g., 
for modifying the setup from vertical to horizontal extraction. 

During preparation of the experiments, blockages in the \isotope[3]{He} cooling circuit 
required two partial warm-ups of the source, 
so that the measurements could start only two days later than initially scheduled.

During operation we observed that, in UCN extractions 
performed with the same accumulation storage times, 
the UCN output decreased rapidly 
by \SI{5\pm1}{\percent} per hour in measurements with vertical extraction. 
In previous and later experiments a clear correlation of the speed of degradation with the 
pressure in the connected extraction guide system was observed 
(e.g., in the low \SI{e-7}{\milli\bar} pressure range, 
a decrease by 46\% has been observed after 24 days of continuous operation).
%
This indicates the rest gas as source of the problem, 
most likely due to a reduction of UCN transmission by adsorption of the gas 
on the cold surfaces of the extraction guide within the cryostat. 
The most probable culprit in the experiments described here is out-gassing 
from the acrylic UCN guide connected to the storage bottle,
despite its NiMo coating on the inside surface.

\subsection{Filling optimization}

\begin{figure}[htb]
\begin{center}
\includegraphics[width=0.5\textwidth]{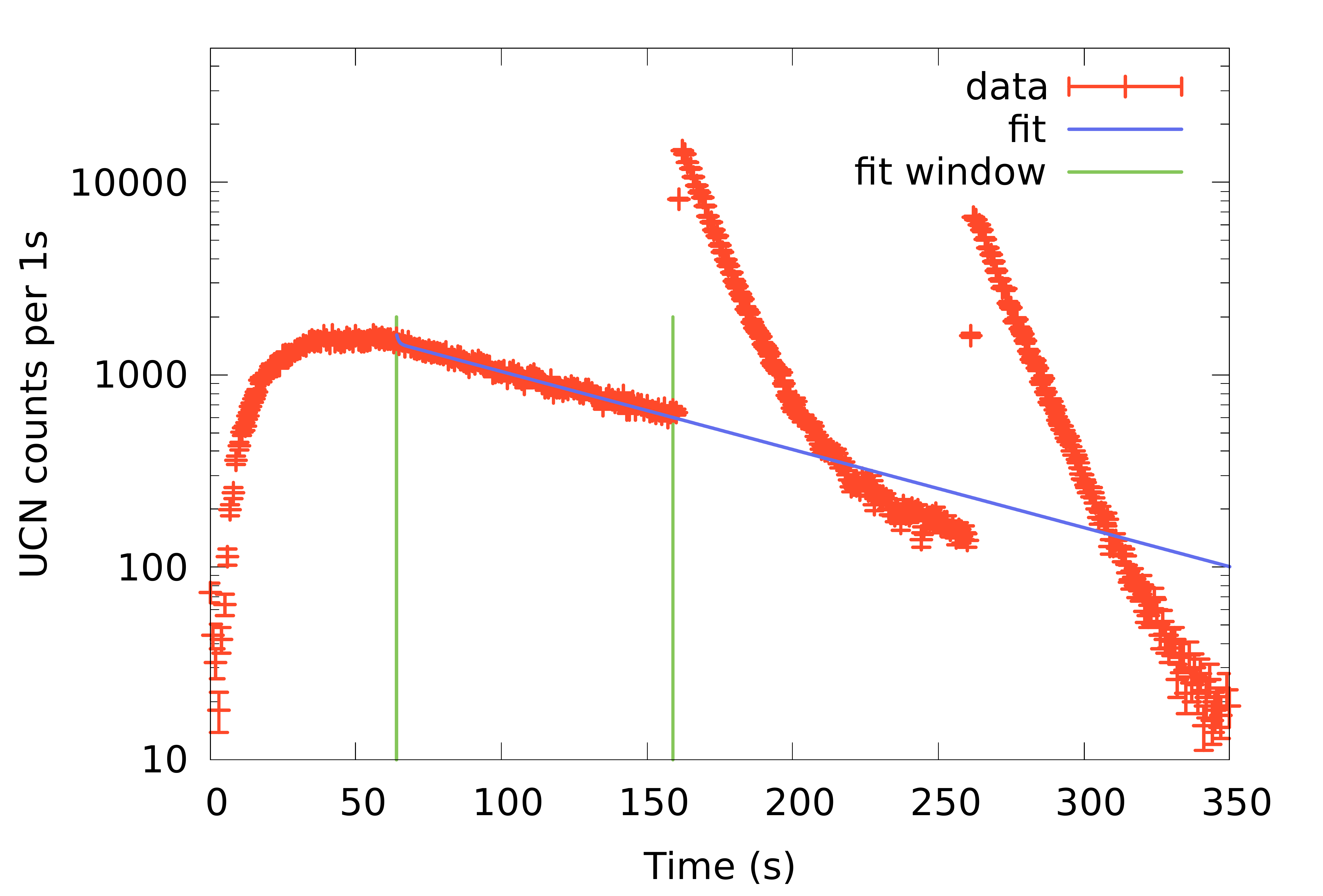}
\end{center}
\caption{
Time spectrum of UCN counts per 1\,s, starting with the opening of the UCN valve (rising count rate) 
which defines time~=~0. 
Shown are measured values for \SI{600}{\second} of accumulation, 
averaged over four UCN extractions with the vertical extraction system behind 
the external storage vessel (two with a storage time of \SI{100}{\second} and two with \SI{200}{\second}, 
respectively, each visible as a separate extraction peak). 
One observes a good saturation of the leakage rate during filling at \SI{50}{\second} 
and its decrease after closure of shutter~1 at the beginning of the storage time, i.e., 
at $t=\SI{60}{\second}$. 
A fit to the leakage rate was performed within the 95\,s long fit window 
indicated by the vertical green lines and extended to the 
duration of the whole period of storage and extraction (blue line).
}
\label{SUN2fillingSpectrum}
\end{figure}

The time spectrum of UCN arrival at the detector is shown in Fig.~\ref{SUN2fillingSpectrum}. 
UCNs were leaking through shutter~2 during the period of filling the external storage vessel, 
and one can observe that after about \SI{50}{\second} the maximum rate was reached. 
From 60 to 160\,s (260\,s for the second peak shown in Fig.~\ref{SUN2fillingSpectrum}) 
UCNs were stored and only UCNs leaking through shutter~2 were observed. 
Then shutter~2 was opened and the UCNs remaining in the storage vessel were detected.
The figure shows a sum of four measurements,
two with a UCN storage time of 100\,s and two with a storage time of 200\,s.

\begin{figure}[htb]
\begin{center}
\includegraphics[width=0.5\textwidth]{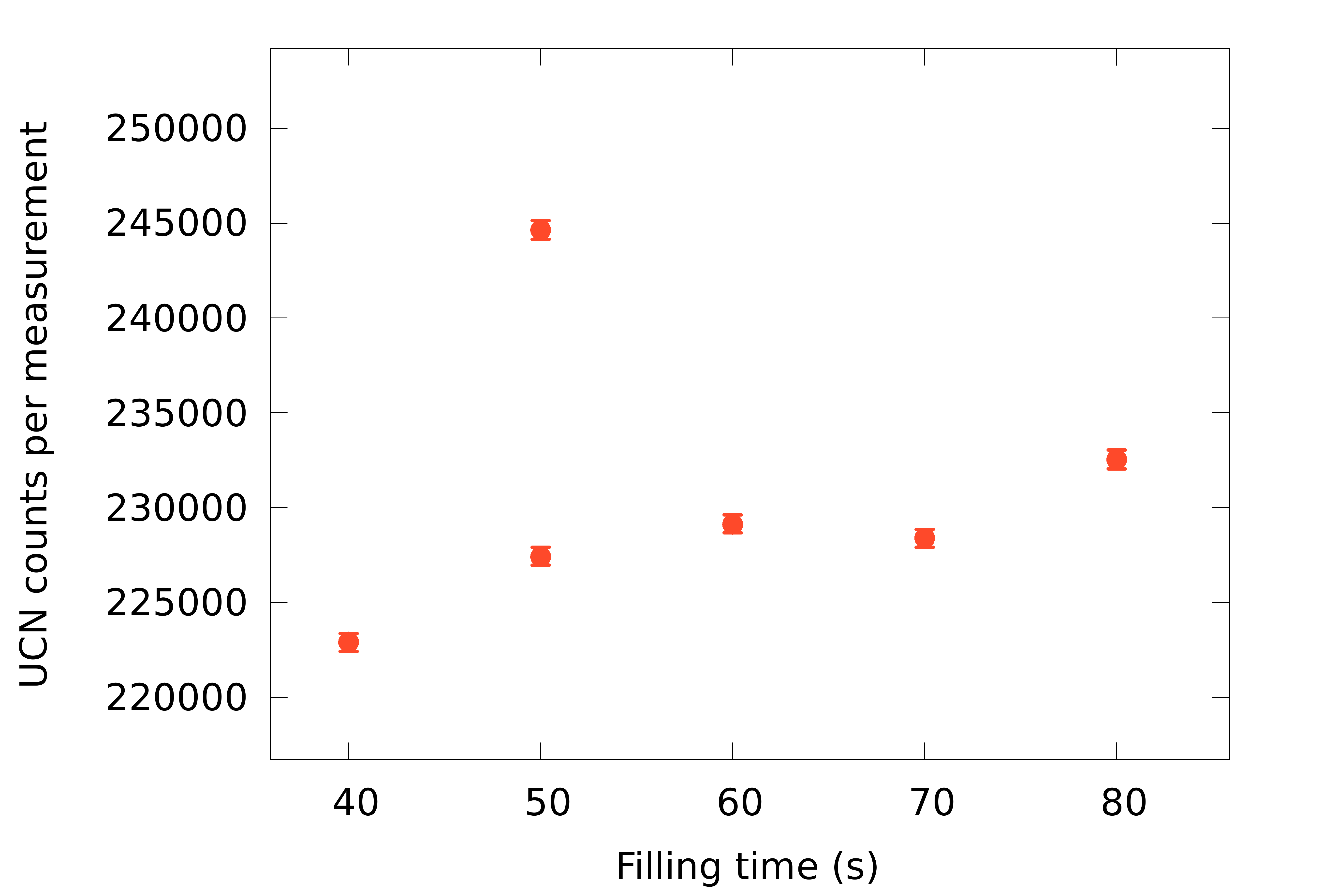}
\end{center}
\caption{
UCN counts per measurement after a storage time of \SI{5}{\second} for various filling times 
of the storage bottle with 600\,s accumulation time of SUN-2 and vertical extraction. 
The two data points at $t=\SI{50}{\second}$ differ because of the decline of UCN source output with time
and also part of the trend of the other data points is due to this effect.
}
\label{SUN2fillingCurve}
\end{figure}

We attempted to optimize the UCN filling time using the procedure described in Sec.~\ref{StandardSequence}, 
in the range from \SI{40}{\second} to \SI{80}{\second} as shown in Fig.~\ref{SUN2fillingCurve}. 
However, as it was quickly observed that the UCN output per accumulation 
was decreasing with time, the scanning of the filling time was aborted after a coarse scan, 
and a filling time of \SI{60}{\second} was chosen. 
The end of the accumulation period was defined as time $t=0$ in the subsequent measurements.

\subsection{Storage measurements}

\begin{figure}[htb]
\begin{center}
\includegraphics[width=0.50\textwidth]{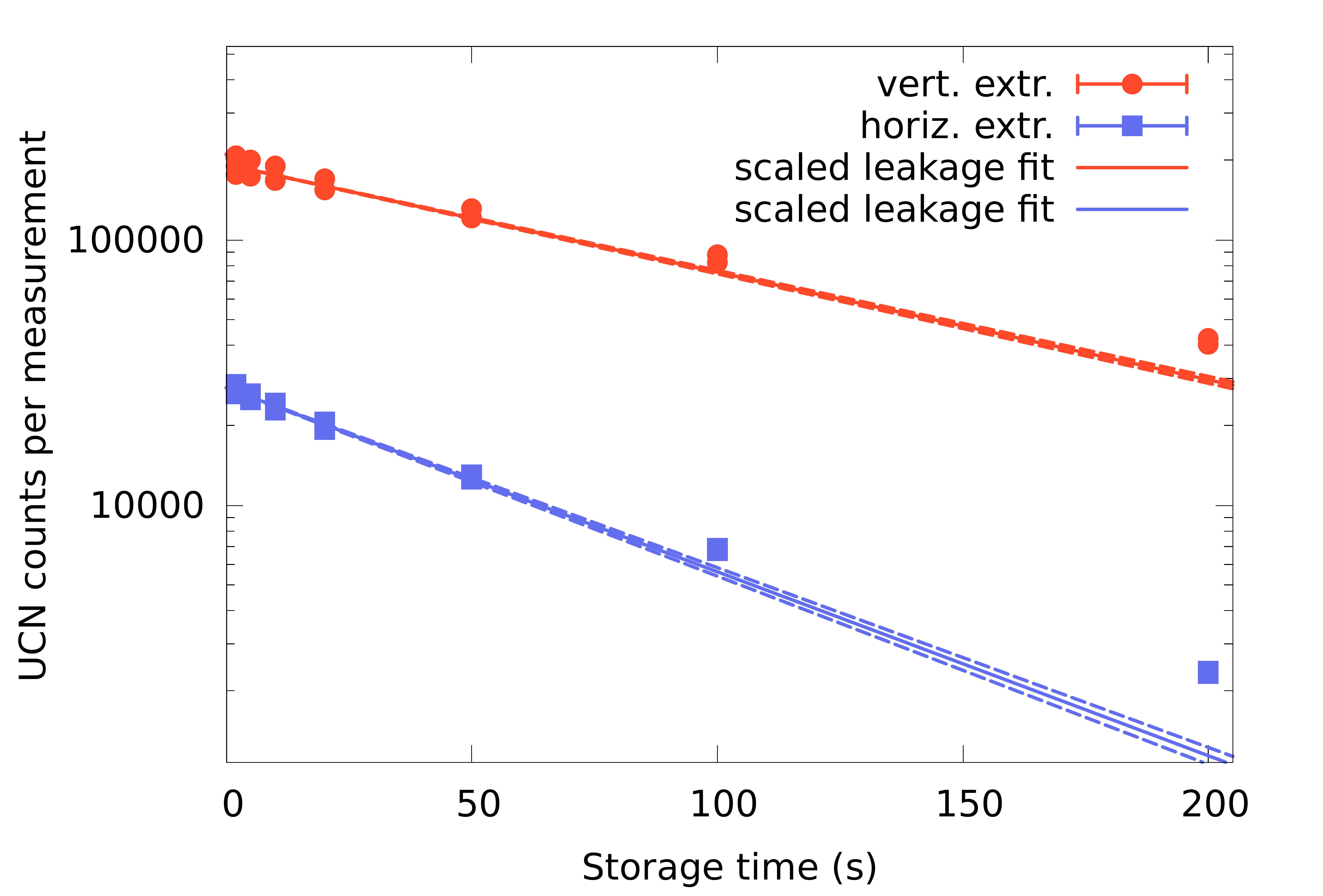}
\end{center}
\caption{
UCN counts per measurement after various storage times for a SUN-2 UCN accumulation time of \SI{600}{\second} 
and vertical and horizontal UCN extractions to the detector. 
The measurement sequence was started with a storage time of \SI{2}{\second} 
and increased up to \SI{200}{\second}. Then one more measurement was done at \SI{2}{\second} 
and afterwards the storage times were decreased from \SI{200}{\second} back to \SI{2}{\second}. 
The differences observed for each pair of measurements are due to the gradual loss of intensity (see text). 
The results of the fits to the storage time constant measured via the shutter leakage rate 
(see Tab.~\ref{SUN2StorageTimeTable})
are indicated by the continuous lines drawn with 1$\sigma$ error bands.
The data point at 200\,s was out of the fit window.
}
\label{SUN2Storage600}
\end{figure}

\begin{figure}[htb]
\begin{center}
\includegraphics[width=0.50\textwidth]{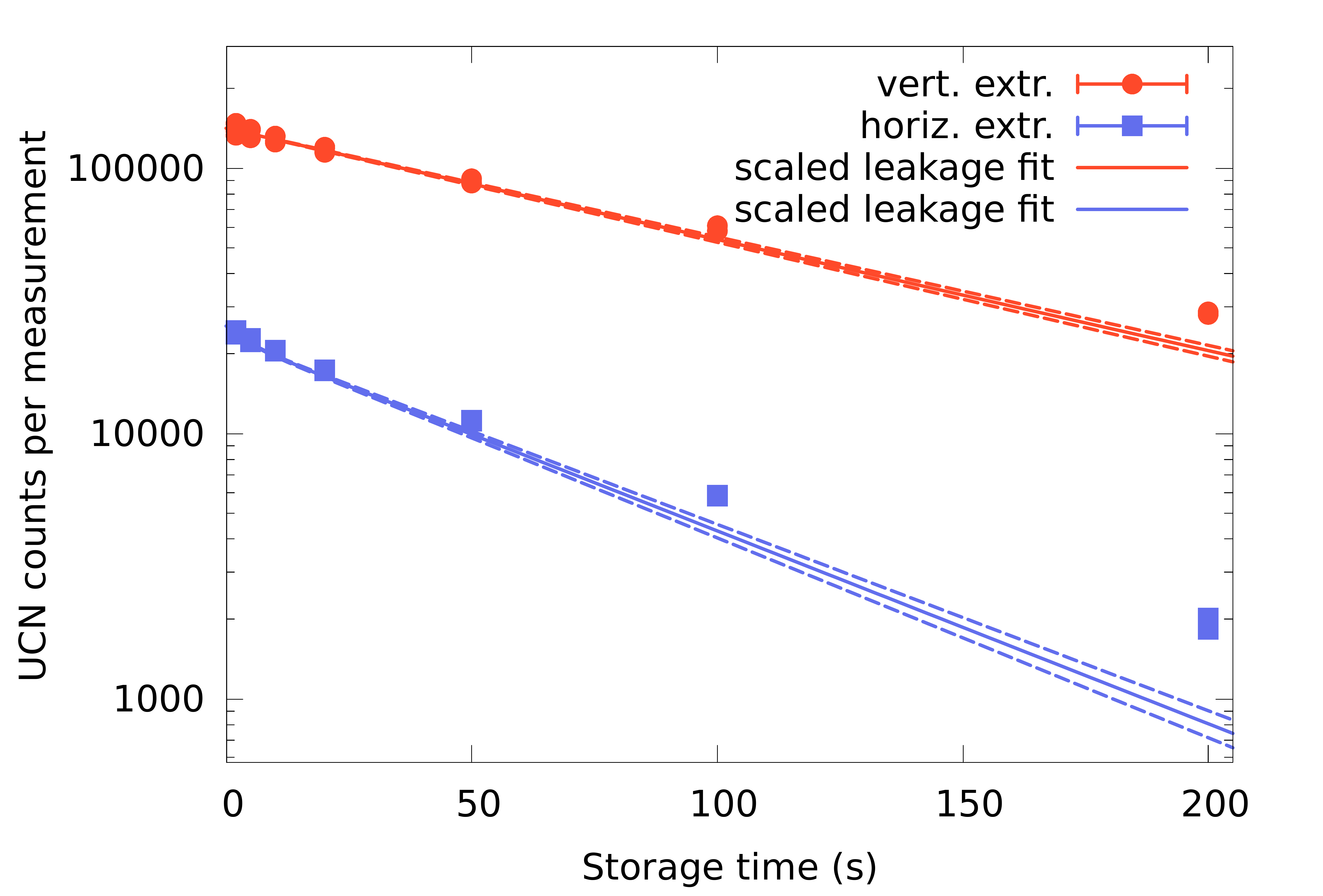}
\end{center}
\caption{
Same as shown in Fig.\ref{SUN2Storage600} but for a SUN-2 UCN accumulation time of \SI{300}{\second}.
}
\label{SUN2Storage300}
\end{figure}

For both UCN source accumulation times, 300 and 600\,s, UCN storage measurements were 
performed with the detector in vertical and horizontal extraction. 
The corresponding results are shown in Fig.~\ref{SUN2Storage600} and Fig.~\ref{SUN2Storage300}.
The amount of UCNs leaking in from the source through shutter~1 
after the end of the filling period 
was determined by a measurement with shutter~1 permanently closed and shutter~2 permanently opened. 
Results are shown in Tab.~\ref{SUN2resultsTable}.

The UCN rate leaking through shutter~2 was fitted in a \SI{95}{\second} 
time window during the storage period, as described in Sec.~\ref{StandardAnalysis}.
In order to determine the storage time constants for each detector set-up and accumulation time, 
the time spectra of four measurements with storage times of \SI{100}{\second} 
and \SI{200}{\second} were added.
In all cases the double exponential fit did either not converge at all or found 
a negligible fast UCN population.
%
Also the double exponential fit was used in order to keep the consistency to the analysis at the other UCN sources.
A single exponential fit seems to be the better model due to the small kinetic energy region 
possible for UCNs between 0 and $\sim$72\,neV, defined by the Fermi potential of Fomblin and the extraction height.
The resulting storage time constants (STC)
are given in Table~\ref{SUN2StorageTimeTable}.
An example of a fit for
vertical extraction and an accumulation of \SI{600}{\second} is shown in Fig.~\ref{SUN2fillingSpectrum}.

\begin{table}[htb]
\begin{center}
\begin{tabular}{c|c|c|c|c|c|c}
	Acc.(s)      & Extr. &$A_1$            & $\tau_1$(s)         & $A_2$            & $\tau_2$(s)        & red.$\chi^2$\\\hline
	\SI{300}{}  & horiz. & \SI{3 \pm 2}{}  & \SI{ 4 \pm 5}{}     & \SI{ 24 \pm 1}{} & \SI{59.7 \pm2.1}{}  & \SI{1.28}{}\\
	\SI{300}{}  & vert.  & ---             & ---                 & \SI{107 \pm 1}{} & \SI{103.7\pm2.2}{}  & \SI{1.32}{}\\
	\SI{600}{}  & horiz. & ---             &  ---                & \SI{ 35 \pm 1}{} & \SI{62.6 \pm1.4}{}  & \SI{1.53}{}\\
	\SI{600}{}  & vert.  & \SI{14\pm 10}{} & \SI{ 1 \pm 1}{}     & \SI{145 \pm 1}{} & \SI{107.0\pm1.7}{}  & \SI{1.34}{}\\
\end{tabular}
\caption{
Parameters resulting from fits to leakage rate. 
In the two cases where the double exponential fit did not converge 
($A_1$ and $\tau_1$ undetermined)
a single exponential fit was used.
}
\label{SUN2StorageTimeTable}
\end{center}
\end{table}

One can see that the relevant storage time constant (STC) for stored UCNs,
$\tau_2$, does only weakly depend 
on the accumulation time of the source. 
The much shorter $\tau_2$ measured in the horizontal extraction 
demonstrates that the spectrum is composed predominantly of low-energy UCNs that can only pass 
the detector Al foil in the vertical extraction setup.

\subsection{UCN Density determination}

Table~\ref{SUN2resultsTable} reports the results for the UCN density measurement 
deduced from the \SI{2}{\second} storage time measurement. 
Listed are the observed UCN counts and resulting densities after storage.
The corresponding UCNs leaking into and out of the bottle 
were subtracted.

\begin{table}[htb]
\centerline{
\begin{tabular}{c|c|c|c|c}
    Acc.   &  Extr.       &     Net UCN          &  Subtracted          & Density       \\
     (s)   &              &     counts           &  leakage counts      & (\SI{}{UCN\per\centi\meter^{3}})   \\ \hline
   \SI{300}{}  & horiz.   &  \SI{24352\pm169}{}  &   \SI{2162\pm46}{}   &   \SI{0.76\pm0.01}{} \\
   \SI{300}{}  & vert.    &  \SI{148132\pm408}{} &   \SI{9066\pm95}{}   &   \SI{4.62\pm0.03}{} \\
   \SI{600}{}  & horiz.   &  \SI{28015\pm181}{}  &   \SI{2409\pm49}{}   &   \SI{0.87\pm0.01}{} \\
   \SI{600}{}  & vert.    &  \SI{207215\pm484}{} &   \SI{13384\pm116}{} &   \SI{6.47\pm0.04}{} \\
   \SI{600}{}  & vert.5   &  \SI{231601\pm495}{} &   \SI{13014\pm116}{} &   \SI{7.23\pm0.04}{} \\
\end{tabular}}
\caption{
SUN-2 results: 
Net UCN counts in \SI{2}{\second} storage measurements, subtracted UCN leakage counts, and determined UCN density.
It doesn't seem undue to take the highest value of Fig.~\ref{SUN2fillingCurve},
``vert.5'',
for the SUN-2 UCN density statement, which is a result of a 5\,s storage measurement,
as the fast degradation of the source performance was caused by the vacuum conditions 
of the setup rather than from the source itself.
}
\label{SUN2resultsTable}
\end{table}

The large difference in UCN densities observed with the horizontal 
and vertical extraction setup 
demonstrates again, that a dominant fraction of UCNs has 
very low kinetic energies.
Therefore, the 0.1\,mm thick
detector Al window has 
a huge effect via Fermi potential and UCN transmission.

Notice the differences of UCN counts for measurements 
performed at same nominal conditions but at different times visible in 
Fig.~\ref{SUN2fillingCurve}, 
which are due to source degradation. 
We point out that no corrections to this were applied in the density determinations as described before. 
By improvement of the poor vacuum conditions prevailing in the connected storage vessel
output degradation might not be an issue.
The source SUN-2 studied here is a prototype for the SuperSUN UCN source 
comprised of a 12\,liter converter vessel with a magnetic 
multipole reflector \cite{Zimmer2015}, which is currently under construction at the ILL.

\section{Measurements at the EDM beamline of the PF2 UCN turbine source at the Institut Laue Langevin}


The ultracold neutron source PF2 of ILL's high-flux reactor 
has been the workhorse of UCN physics in the last three decades.
It is based on the so-called 'Steyerl' turbine 
which mechanically Doppler-shifts neutrons towards UCN velocities.
The facility is described in detail in~\cite{Steyerl1986}.
Stored UCN densities of \SI{36}{UCN\per\centi\meter^3} are reported.

PF2 offers four different UCN beamlines, 
each having slightly different fluxes and energy spectra 
due to the different extraction geometries.
Three of these beamlines are operated in a time-sharing mode. 
All measurements described here were performed at the EDM beamline 
which is known to have the highest 
UCN flux.

\subsection{Setup at the EDM beamline}

\begin{figure}[htb]
\begin{center}
\includegraphics[width=0.5\textwidth]{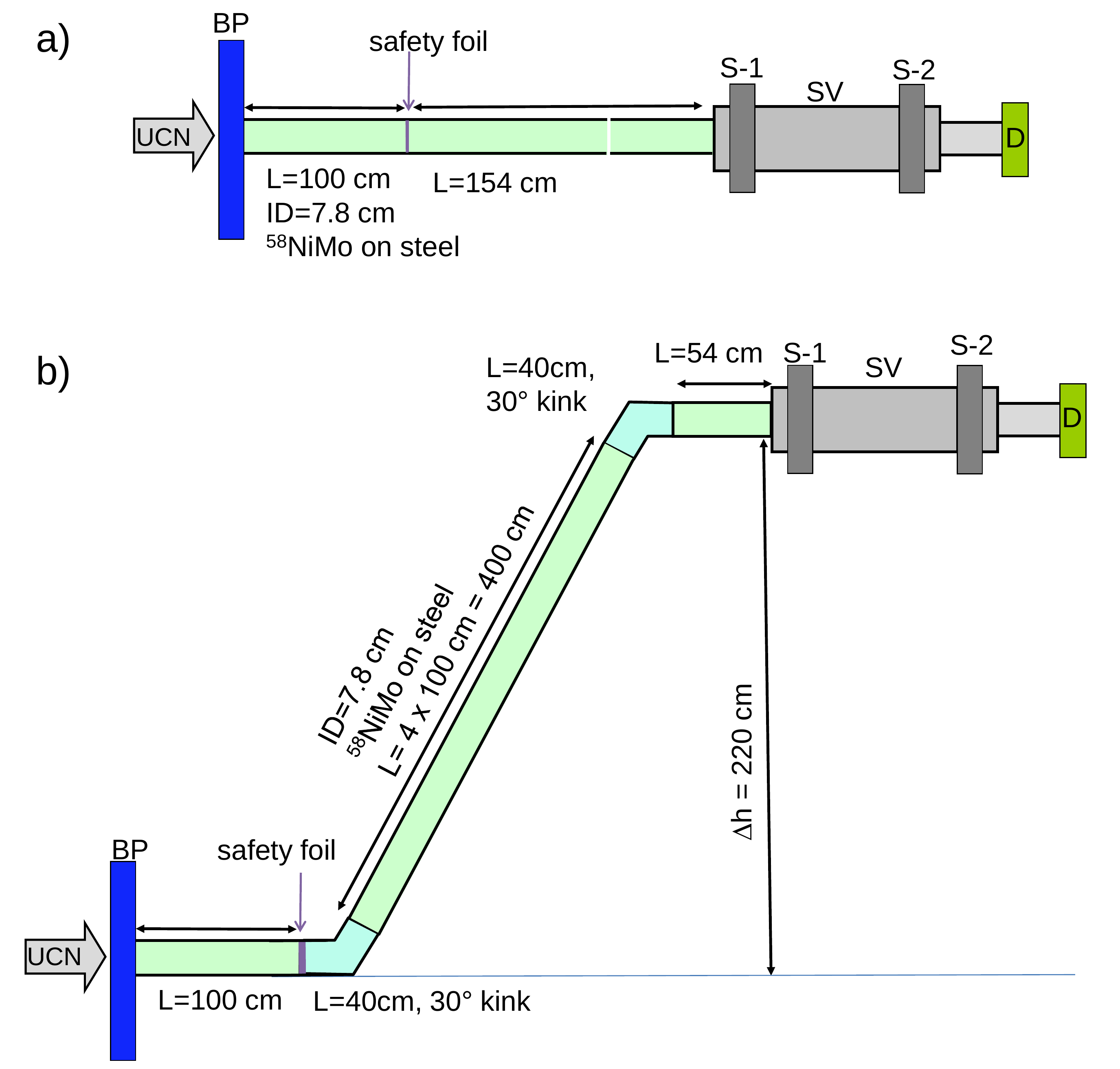}
\end{center}
\caption{
Sketch of the setup at PF-2 (not to scale): 
a) at the height of the turbine exit and 
b) on the EDM platform. 
Components as described in Fig.~\ref{SUN2setup}. 
In addition, the location of the safety foil after the beamport is indicated.
}
\label{PF2setup}
\end{figure}

In the frame of this work, all UCN density measurements 
were performed in two principle configurations: 
First, the bottle was connected to the beamport using 
\SI{2.54}{\meter}
of straight horizontal guide tubes. 
Second, to shift the UCN spectrum to lower velocities due to gravity, 
the setup was installed approximately 
\SI{2.2}{\meter} 
above the turbine exit 
using additional \SI{4}{\meter} of guides.
At this second position, the former RAL-Sussex-ILL nEDM 
experiment~\cite{Baker2006} was located.
All UCN guides used in the measurements at PF2 were stainless-steel guides with an 
inner diameter of
\SI{78}{\milli\meter}, coated with 
$^{58}$NiMo 
on the inside. 
The two installations are sketched in
Fig.~\ref{PF2setup}.

\begin{figure}[ht]
\begin{center}
\includegraphics[width=0.50\textwidth]{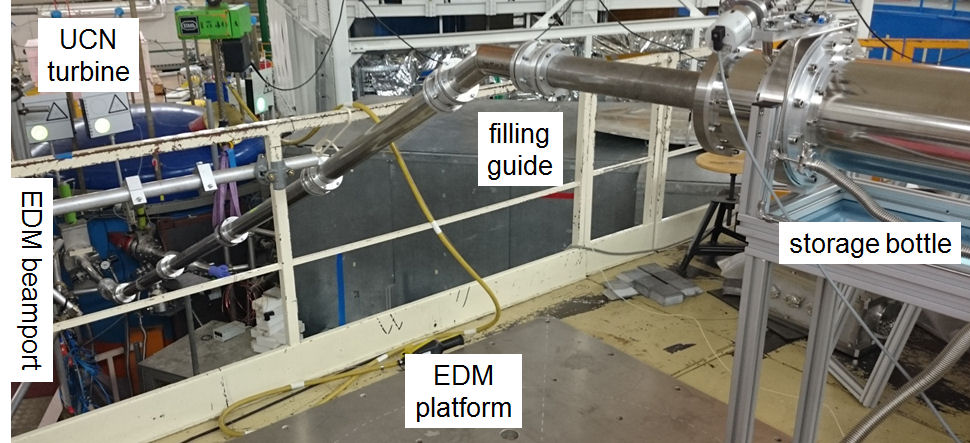}
\caption{PF2 EDM platform: 
Storage bottle setup 
with the long UCN guiding section
to the EDM platform at PF2. 
The blue UCN turbine is visible 
on the left side in the back.
}
\label{PF2PlatformPhotos}
\end{center}
\end{figure}

For safety reasons at PF2, an aluminium foil 
(AlMg3, \SI{100}{\micro\meter} thick) 
separates the vacuum of the experiment from the vacuum of the turbine. 
Within this campaign, all measurements were performed with and without this safety foil. 
To gain additional information on the spectra of the stored UCNs, 
all measurements were repeated with horizontal and vertical extraction.
A photo of the beamline setup is shown in Fig.~\ref{PF2PlatformPhotos}.

\subsection{Operating conditions of the UCN source}
\label{PF2Conditions}

The described measurements were performed in two different reactor cycles in 2015. 
The measurements directly at the turbine exit were performed at a nominal reactor power 
of \SI{52.5\pm0.3}{\mega\watt}. 
The other measurements on top of the nEDM platform took place during the first 
two operating days of the next cycle, and the reactor was still ramping to its final power. 
Here, the two measurements without vacuum safety foil were taken at a reactor power of 
about \SI{49.0\pm0.5}{\mega\watt}. 
The measurements with vacuum safety foil were taken at a slightly higher 
reactor power of about \SI{50.8\pm0.5}{\mega\watt}.
The count rates given below were not corrected for the different reactor powers.

The pressure inside the storage bottle was below \SI{2e-4}{\milli\bar}.
During all measurements, all UCN beamlines were used in time-sharing-mode. 
To fill the storage bottle, the UCN beam was requested for a certain time 
using a dedicated PF2 control software. 
Then, a dedicated guide switching device moves an extraction guide towards the PF2/EDM-position. 
After filling the bottle, it can be released and other UCN positions can be served.
If this is not the case, the UCN intensity on the first shutter of the storage bottle stays high 
even after filling the bottle is finished. 
The storage bottle shutters are not completely UCN tight. 
Therefore, relatively large amounts of UCNs leak into the bottle during storage.

To have identical background conditions for different measurement runs, 
it was chosen to always keep the beamline switch at the EDM beamline 
until the counting time of the measurement was finished. 
The leaking UCNs have a minor effect on the measured UCN density after 
\SI{2}{\second}, 
but need to be taken into account to determine the storage time constant.

\subsection{Filling optimization}

\begin{figure}[ht]
\begin{center}
\includegraphics[width=0.50\textwidth]{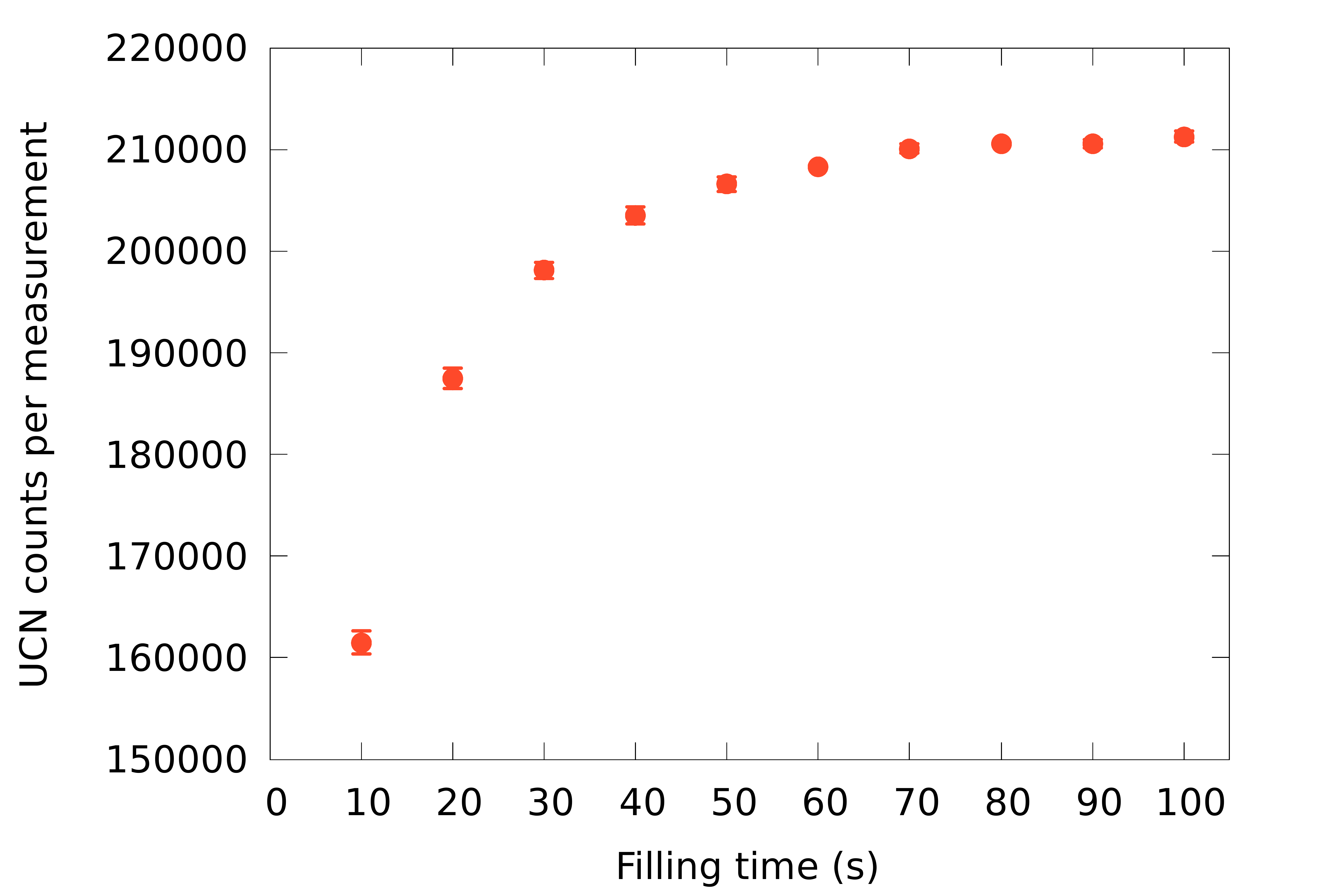}
\end{center}
\caption{
UCN counts per measurement 
after \SI{5}{\second} of storage using various filling times
measured at turbine exit height
with vertical extraction and with the vacuum separation foil in place. 
}
\label{PF2fillingCurve1}
\end{figure}

\begin{figure}[ht]
\begin{center}
\includegraphics[width=0.50\textwidth]{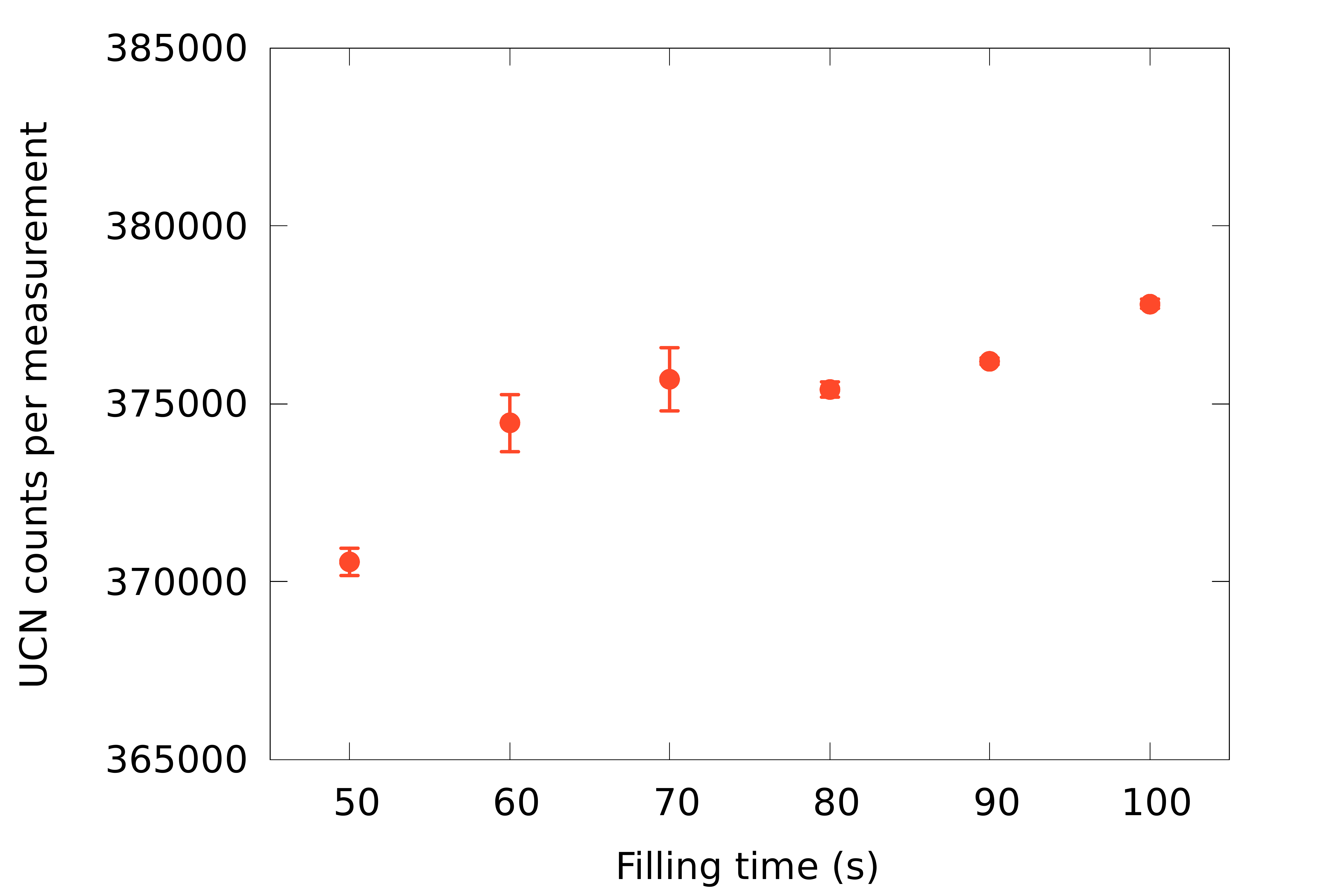}
\end{center}
\caption{Same as in Fig.~\ref{PF2fillingCurve1}
but measured on top
of the EDM platform with vertical extraction, without vacuum separation foil.}
\label{PF2fillingCurve2}
\end{figure}

In both configurations, on top of the platform and at turbine exit height, 
the filling time was optimized as described in Sec.~\ref{StandardSequence}. 
Time zero was defined by the turbine signal 
indicating that the EDM port is served with UCNs.
The results are shown in Figs.~\ref{PF2fillingCurve1}
and~\ref{PF2fillingCurve2}. 
A filling time of \SI{70}{\second} 
was chosen for the measurements, to be long enough to fit to both positions
and conditions.

\subsection{Storage measurements}

\begin{figure}[ht]
\begin{center}
\includegraphics[width=0.5\textwidth]{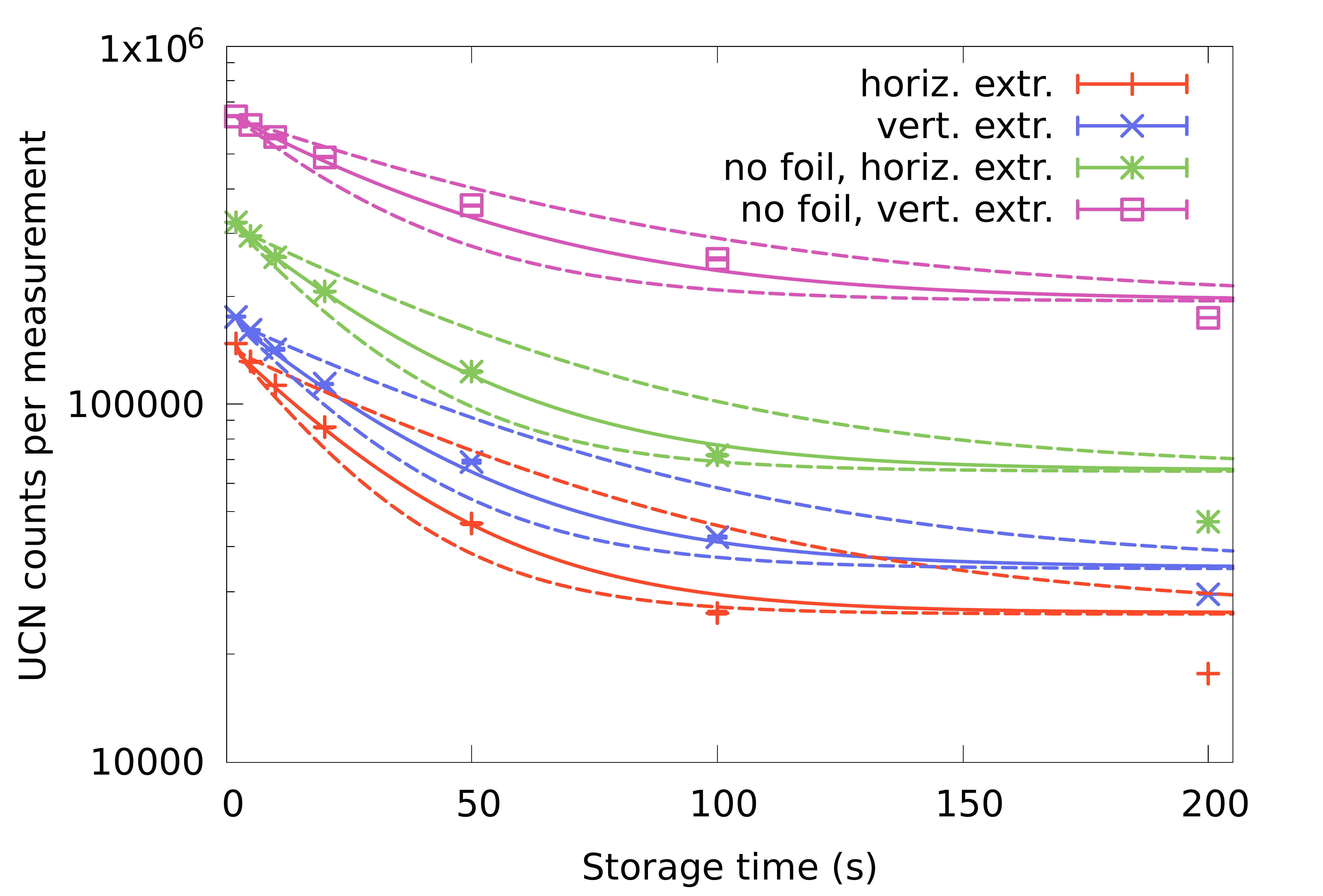}
\caption{
UCN counts per measurement
after various storage times 
measured at turbine exit height. 
The results of the fits to the leakage rates are indicated
by the continuous lines drawn with hardly visible 1$\sigma$ error bands
for the determined storage times of 
27.6, 32.0, 32.1, and 41.8\,s, respectively,
see Tab.~\ref{PF2DOWNStorageTimeTable}.
The data point at 200\,s was out of the fit window.
Error bands are smaller at later times, as the UCN counts are
then dominated by leakage.
UCN count rates are higher for
both measurements without aluminum safety foil than with foil.
}
\label{PF2StorageDownstairs}
\end{center}
\end{figure}

\begin{figure}[ht]
\begin{center}
\includegraphics[width=0.5\textwidth]{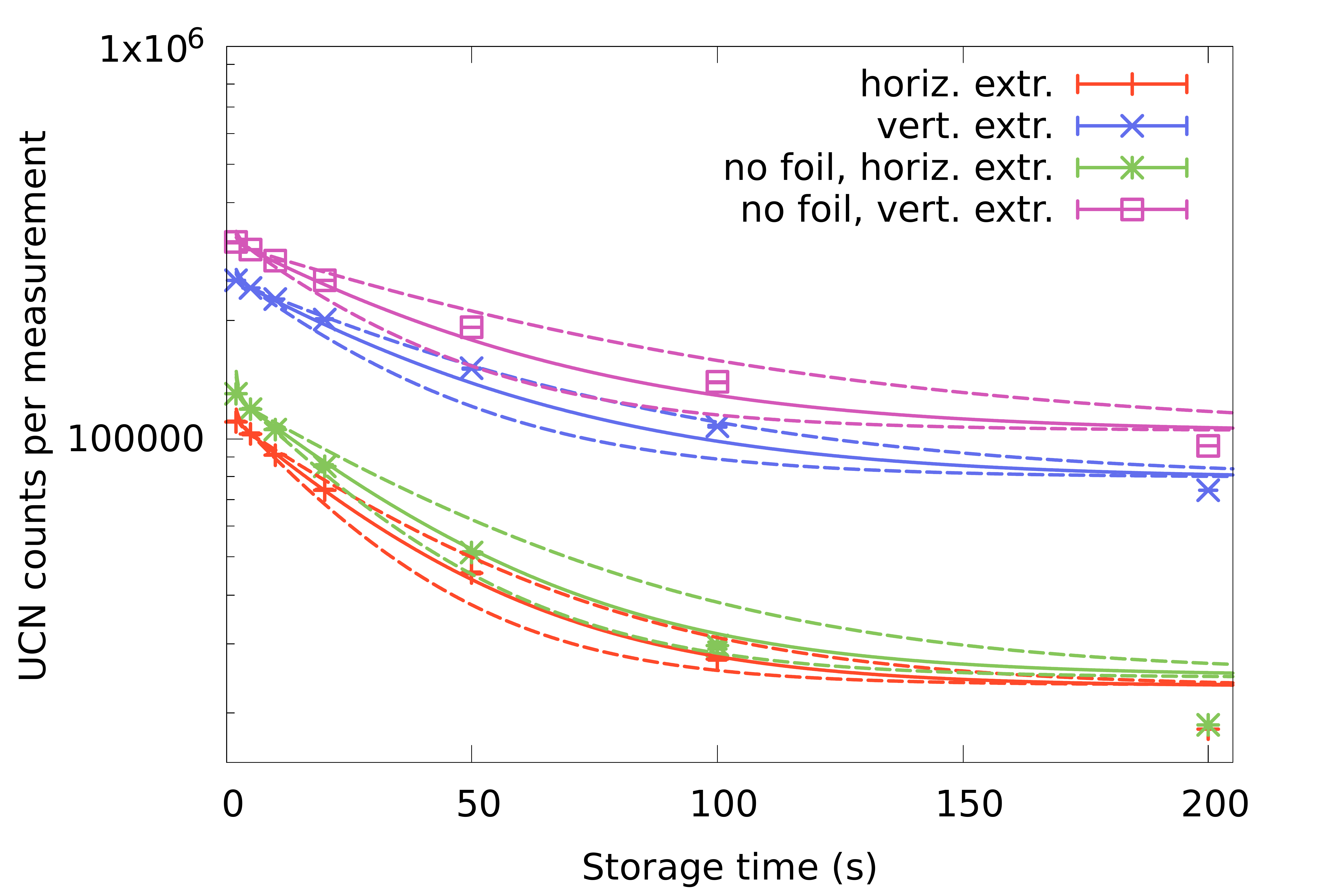}
\caption{
Same as shown in Fig.~\ref{PF2StorageDownstairs},
but measured on the EDM platform. 
The results of the fits to the leakage rates are indicated
by the continuous lines drawn with hardly visible 1$\sigma$ error bands
for the determined storage times of 
33.1, 44.8, 36.5, and 45.9\,s, respectively,
see Tab.~\ref{PF2PLATStorageTimeTable}.
The data point at 200\,s was out of the fit window.
Error bands are smaller at later times, as the UCN counts are
then dominated by leakage.
Measurements with and without aluminum foil are interspersed.}
\label{PF2StoragePlatform}
\end{center}
\end{figure}

Storage measurements were performed
in both configurations,
with horizontal and vertical extraction, and
with and without vacuum safety foil.
The results, UCN counts as a function of storage time, are shown in
Fig.~\ref{PF2StorageDownstairs} and 
Fig.~\ref{PF2StoragePlatform}, respectively.

It is worth noting that the lifting of the experiment 
by \SI{2.2}{\meter} in height 
lowers the average kinetic energy of UCNs stored in the bottle due to gravity. 
As a consequence, a signiﬁcant fraction of the stored UCNs is too slow 
to overcome the Fermi potential of the detector entrance window 
(\SI{54}{\nano\electronvolt}),
and thus can only be detected with the detector mounted 
\SI{1}{\meter} lower.

\begin{figure}[htb]
\begin{center}
\includegraphics[width=0.50\textwidth]{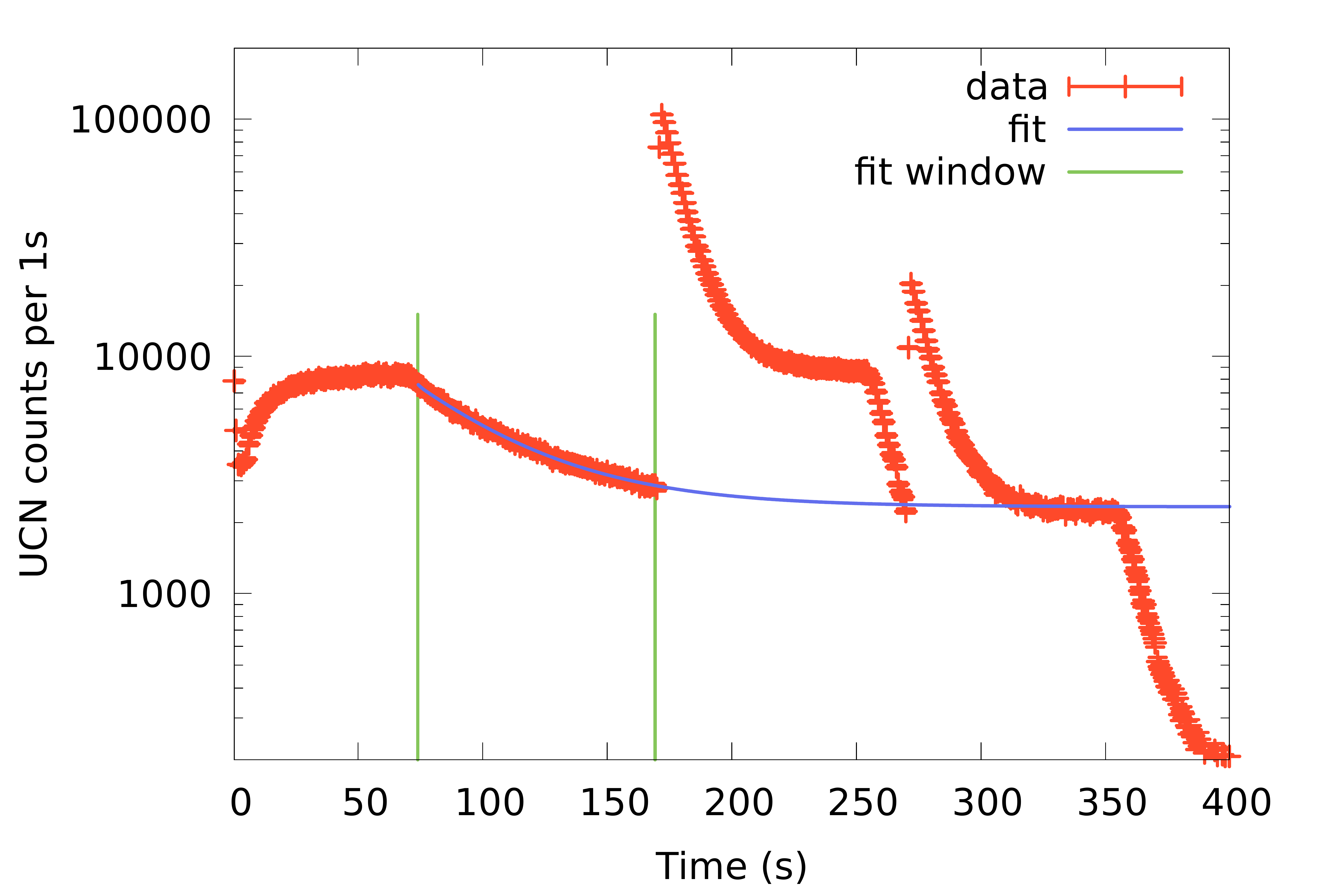}
\caption{
Time spectrum of UCN counts in the detector, measured with vertical extraction and no 
safety foil at turbine exit height. Sum of four measurements with \SI{100}{\second}
storage time and one measurement with \SI{200}{\second} storage time.
Starting from a very low UCN rate when the turbine's extraction tube is at a different beam line position, 
the rate of UCNs leaking through shutter~2 during filling rises 
and saturates after about \SI{50}{\second}.
After shutter~1 is closed at $t=\SI{70}{\second}$, 
the UCN rate drops for the storage time
until the first emptying peak at $t=\SI{170}{\second}$. A second emptying
peak (\SI{200}{\second} storage time) can be seen at $t=\SI{270}{\second}$.
Afterwards, the count rate stabilizes at the high leakage rate of 
about \SI{2}{\kilo\hertz}
until the turbine extraction tube moves away at $t=\SI{350}{\second}$.
The fit to the leakage rate (blue line) of UCNs
during these long storage measurements 
is indicated, the vertical green lines show the fit window.
}
\label{PF2example20}
\end{center}
\end{figure}

A typical time spectrum of UCN counts 
in a storage measurement is shown in Fig.~\ref{PF2example20}.
All time spectra are qualitatively similar, and differ only 
in the neutrons leaking through with closed shutters, 
and in the size of the emptying peak.

As the PF2 turbine delivers a constant flux of UCNs, 
the rate impinging on shutter~1 
was very high during the storage and counting times. 
As the entrance shutter of the storage volume is not completely UCN-tight,
UCNs were steadily leaking in during storage and counting. 
This effect needs to be taken into account for the determination 
of the storage time constant. 
Therefore, a modification of the standard fit function
[Eqn.~(\ref{standardEquation})] was necessary. 
An extension of 
Eqn.~(\ref{standardEquation})
with a constant background term corresponding to the leakage 
cannot describe the measurements with longer storage times, 
where a time dependence of leaking UCNs, in and out, becomes relevant. 
Therefore, a background term was introduced, which scales with a time $\tau_1$ 
and the fit function reads as: 
\begin{equation}
 N(t)\propto R(t) = A_1\times e^{-t/\tau_1} + A_2\times e^{-t/\tau_2} + L\times\tau_1
\end{equation}
where $N(t)$ is the total amount of UCNs inside the storage bottle, 
$R(t)$ the rate of UCNs leaking through shutter~2 as measured in the detector, 
$A_1$ and $A_2$ are population constants, and 
$\tau_1$ and $\tau_2$ exponential decay time constants.
The leakage rate 
$L$ describes the UCNs leaking in per second 
through shutter~1.
It was measured for every setup 
and is given in Table~\ref{PF2resultsTableDown}
and Table~\ref{PF2resultsTablePlat}. 

\begin{table}[htb]
\begin{center}
    \begin{tabular}{c|c|c|c|c|c|c}
	    Foil & Extraction &$A_1$  & $\tau_1$ (s) & $A_2$ & $\tau_2$ (s) & red.$\chi^2$\\\hline
	    Yes  & horiz. & \SI{13 \pm 10}{} & \SI{ 0.6 \pm0.1}{} & \SI{186 \pm 2}{} & {28$^{+28}_{-6}$}{} &\SI{1.32}{}\\
	    Yes  & vert.  & \SI{11 \pm 10}{} & \SI{ 0.7\pm0.1}{}  & \SI{198 \pm 2}{} & {32$^{+24}_{-7}$}{} &\SI{1.26}{}\\
	    No   & horiz. & \SI{17 \pm 14}{} & \SI{ 0.7 \pm0.1}{} & \SI{380 \pm 3}{} & {32$^{+19}_{-8}$}{} &\SI{1.35}{}\\
	    No   & vert.  & \SI{10 \pm 14}{} & \SI{ 1.1 \pm0.1}{} & \SI{519 \pm 4}{} & {42$^{+23}_{-13}$}{} &\SI{1.26}{}\\
    \end{tabular}
    \caption{Parameters resulting from fits to leakage rate at turbine exit height.}
    \label{PF2DOWNStorageTimeTable}
\end{center}
\end{table}

Using this model, 
storage time constants were then derived from storage measurements 
longer than \SI{100}{\second}
with a $\chi^2$
between \SI{1.25}{} and \SI{1.35}{}, which is comparable to the $\chi^2$ values
obtained for the corresponding fits without leakage term evaluated 
at all other (pulsed) UCN sources. 
The results
are given in Table~\ref{PF2DOWNStorageTimeTable} and 
Table~\ref{PF2PLATStorageTimeTable}.
$\tau_1$ in Tab.~\ref{PF2DOWNStorageTimeTable} and
Tab.~\ref{PF2PLATStorageTimeTable}
can be better interpreted as a 'background correction'
to the overall UCN counts than as a
storage time representing a physical UCN population.
However, there is a strong correlation between the
result for $\tau_2$ and the large leakage term $L\times\tau_1$
which cannot be directly measured in the used setup.
We estimated the uncertainty of the storage time constant $\tau_2$
due to the huge correlation with the leakage term
to be 25\%.
This result was verified in in two independent analyses
using varying fit conditions.
This added a significant asymmetric uncertainty to $\tau_2$.
For a more detailed understanding of the storage time constants and 
leakage rates one would have
to take a full UCN velocity spectrum into account.
This was beyond the scope of the present work, 
as its main goal, 
the UCN density measurement, is hardly affected.

\begin{table}[htb]
\begin{center}
    \begin{tabular}{c|c|c|c|c|c|c}
	    Foil & Extraction &$A_1$  & $\tau_1$ (s) & $A_2$ & $\tau_2$ (s) & red.$\chi^2$\\\hline
	    Yes   & horiz. & \SI{8 \pm 7}{}   & \SI{0.6 \pm0.1}{} & \SI{102 \pm 1}{} &  {33$^{+8}_{-7}$}{}  & \SI{1.27}{}\\
	    Yes   & vert.  & \SI{12 \pm 8}{}  & \SI{1.0 \pm0.1}{} & \SI{158 \pm 2}{} &  {45$^{+13}_{-12}$}{}  & \SI{1.25}{}\\
	    No    & horiz. & \SI{25 \pm 9}{}  & \SI{0.5 \pm0.1}{} & \SI{122 \pm 1}{} &  {36$^{+12}_{-7}$}{}  & \SI{1.31}{} \\
	    No    & vert.  & \SI{15 \pm 9}{}  & \SI{1.0 \pm0.1}{} & \SI{189 \pm 3}{} &  {46$^{+28}_{-14}$}{}  & \SI{1.24}{} \\
    \end{tabular}
    \caption{Parameters resulting from fits to leakage rate on the EDM platform.}
    \label{PF2PLATStorageTimeTable}
\end{center}
\end{table}

\subsection{UCN density determination}

Using the UCN counts from the storage 
measurements with a storage time of \SI{2}{\second},
the UCN densities 
were determined.
Measured UCN counts, subtracted leakage, and densities at turbine height 
are listed in Table~\ref{PF2resultsTableDown}, the corresponding data for
the measurements on the EDM platform are listed in Table~\ref{PF2resultsTablePlat}.

\begin{table}[htb]
\begin{center}
\begin{tabular}{c|c|c|c|c|c}
   Foil & Extr.   &   Net UCN            & Leakage        &    Subtracted       &   Density  \\
        &         &   counts             & rate (Hz)      &    leakage counts   &   (\SI{}{UCN\per\centi\meter^{3}})   \\ \hline
   Yes  & horiz.  & \SI{147740\pm232}{}  & \SI{750\pm8}{} &  \SI{60000}{} & \SI{4.61\pm0.02}{} \\
   Yes  & vert.   & \SI{175459\pm444}{}  & \SI{760\pm9}{} &  \SI{60800}{} & \SI{5.48\pm0.03}{} \\
   No   & horiz.  & \SI{321730\pm777}{}  & \SI{1400\pm12}{}& \SI{112000}{}& \SI{10.04\pm0.06}{} \\
   No   & vert.   & \SI{637795\pm1166}{} & \SI{2050\pm9}{} & \SI{164000}{}& \SI{19.90\pm0.11}{} \\
\end{tabular}
\caption{
PF2 results at turbine exit height: 
Net UCN counts in \SI{2}{\second} storage measurements,
leakage rate $L$,
subtracted UCN leakage counts, 
and determined UCN density.
}
\label{PF2resultsTableDown}
\end{center}
\end{table}

\begin{table}[htb]
\begin{center}
\begin{tabular}{c|c|c|c|c|c}
   Foil & Extr.   &   Net UCN           & Leakage       &   Subtracted &   Density  \\
        &         &   counts            & rate (Hz)     &   leakage counts   &   (\SI{}{UCN\per\centi\meter^{3}})   \\ \hline
   Yes  & horiz.  & \SI{110636\pm497}{} & \SI{434\pm5}{} &  \SI{34720}{} &  \SI{3.45\pm0.02}{} \\
   Yes  & vert.   & \SI{253502\pm392}{} & \SI{750\pm7}{} &  \SI{60000}{} &  \SI{7.91\pm0.04}{} \\
   No   & horiz.  & \SI{130182\pm213}{} & \SI{555\pm17}{} &  \SI{44400}{} & \SI{4.06\pm0.02}{} \\
   No   & vert.   & \SI{316709\pm944}{} & \SI{904\pm12}{} &  \SI{72320}{} & \SI{9.88\pm0.06}{} \\
\end{tabular}
\caption{
PF2 results on the EDM platform: 
Net UCN counts in \SI{2}{\second} storage measurements,
leakage rate $L$,
subtracted UCN leakage counts, 
and determined UCN density.		
}
\label{PF2resultsTablePlat}
\end{center}
\end{table}

The following conclusions can be derived from the measurements at PF2:
At the EDM beamline 
a relevant UCN fraction is below 54\,neV,
and does not pass the Al safety foil.
Lowering the detector by 1\,m increases the UCN counts.
Optimizing the UCN transmission of the vacuum 
safety foil is important.


\section{Measurements at the solid D$_2$ source at the TRIGA reactor of the University of Mainz}
\label{sec:TRIGA}


At the TRIGA reactor of 
the Johannes Gutenberg University Mainz, Germany, 
measurements were
performed in November 2015.
The UCN source at the radial beamtube~D is described in detail in~\cite{Lauer2013,Karch2014},
where measured UCN densities of up to \SI{25}{UCN\per\centi\meter^{3}}
were reported in smaller storage bottles.
%
%
Storage measurements were performed with this source before 
its upgrade became operational in September 2016.

The TRIGA reactor was used in a pulsed mode for UCN production, with a pulse-length 
of about \SI{30}{\milli\second}. 
Pulses of about \SI{9.5}{\mega\watt\second} were produced every \SI{\sim12}{\minute}.

From the solid D$_2$ converter, UCNs are extracted horizontally using a stainless-steel
UCN guide with an inner diameter of \SI{66}{\milli\meter}. 
In solid deuterium UCNs experience a 
horizontal boost in kinetic energy of \SI{105}{\nano\electronvolt} as 
observed in Mainz for the first time~\cite{Altarev2008}.
There is no height compensation inside the source,
hence the UCN spectrum starts at this energy.

\subsection{Setup at the TRIGA}

Beamtube~D is horizontally directed straight at the reactor core. 
The in-pile cryostat was 3\,cm from its foremost position as in Ref.~\cite{Karch2014}.
Outside the reactor shielding, the source
vacuum is separated from the experiment vacuum by an AlMg$_3$ foil of 
\SI{100}{\micro\meter} thickness.

\begin{figure}[htb]
\begin{center}
\includegraphics[width=0.5\textwidth]{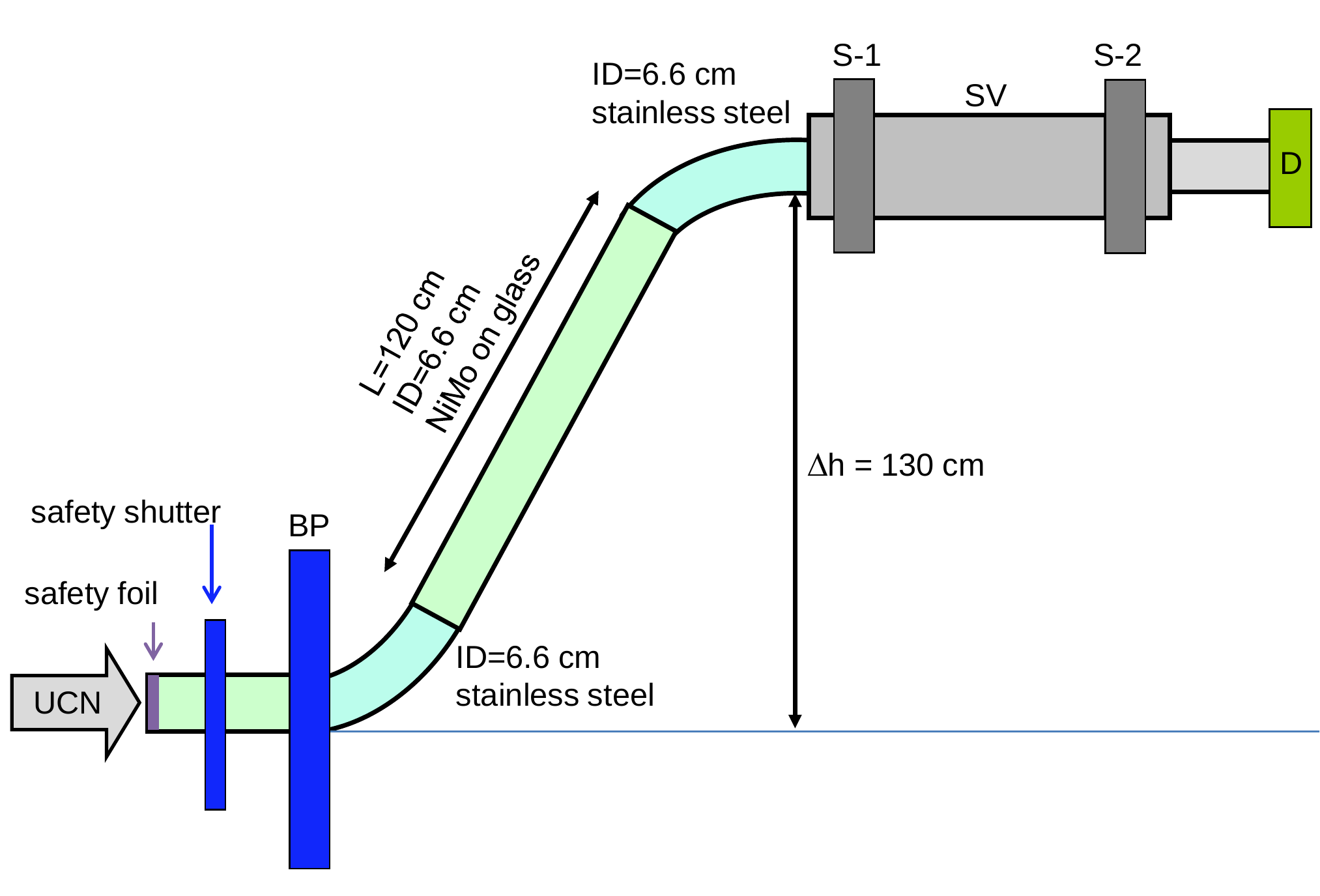}
\end{center}
\caption{
Sketch of the setup at TRIGA Mainz (not to scale): 
Components as described in Fig.~\ref{SUN2setup}.
In addition the location of the safety foil is indicated which is 
about 1\,m upstream from the beamport.
}
\label{TRIGAsetup}
\end{figure}

\begin{figure}[htb]
\begin{center}
\includegraphics[width=0.5\textwidth]{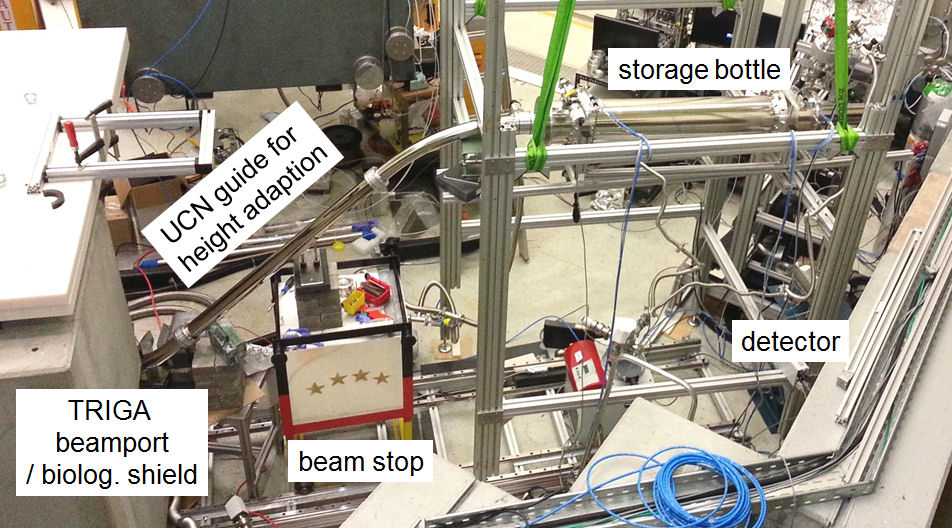}
\caption{
Photo of the storage setup 
with vertical extraction
at beamtube~D
of the TRIGA Mainz reactor. 
}
\label{TRIGAPhotos}
\end{center}
\end{figure}

The storage bottle was connected to the beamport 
via an S-shaped beamline tube
made from two stainless-steel bends of \SI{800}{\milli\meter} bending radius
and an outer diameter of \SI{70}{\milli\meter}, 
with straight UCN guides of various lengths in-between
(see sketch in Fig.~\ref{TRIGAsetup}). 
Between the UCN port and the S-shaped beamline, an additional fast
UCN shutter was installed. 
It was kept open during all density measurements
except for the leakage measurements.
A photo of the setup
is shown in Fig.~\ref{TRIGAPhotos}.

\begin{figure}[htb]
\begin{center}
\includegraphics[width=0.5\textwidth]{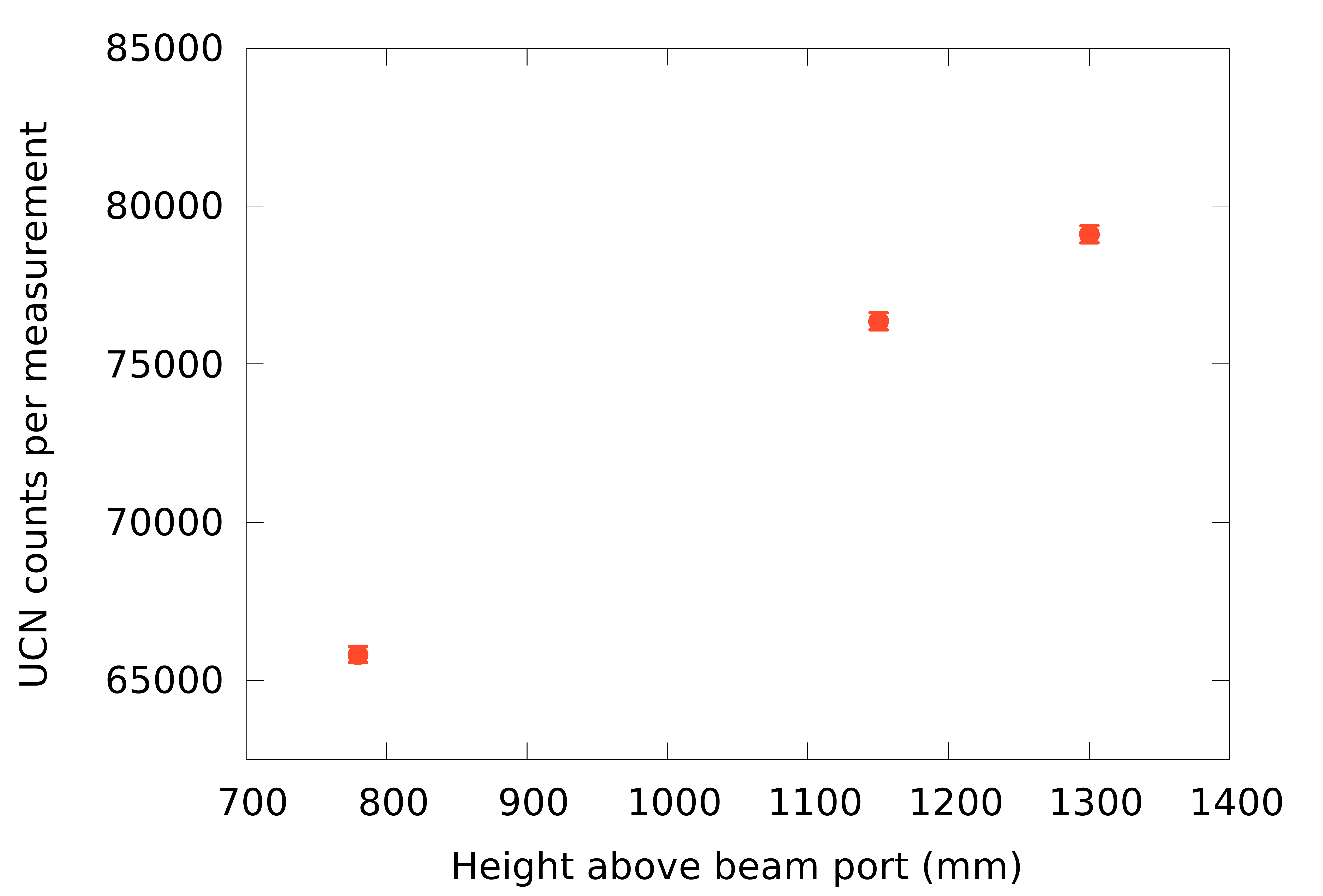}
\caption{
UCN counts per measurement 
after \SI{2}{\second} of storage for mounting positions of the
storage bottle at various heights above the beamport.
The filling times were optimized for each height.
Due to space restrictions the setup could not be mounted
at larger heights.
}
\label{TRIGAHeigthOpt}
\end{center}
\end{figure}

The height of the installation was varied 
in order to find the maximal UCN counts and density.
Fig.~\ref{TRIGAHeigthOpt} shows the counts 
measured after \SI{2}{\second} of storage for various heights
with the maximum at
\SI{130}{\centi\meter} above the beam port. 
At this position
a \SI{120}{\centi\meter} long
glass guide was installed between the two bends
for height adaption.

\subsection{Operating conditions during the measurements}

The average pulse energy of the reactor pulses 
with an insertion of an excess reactivity of about 2\textdollar~ was
\SI{9.5\pm0.1}{\mega\watt\second} during these measurements. 
The amount of deuterium in the source was 
8\,moles.
\isotope{H_2} was used as a pre-moderator. 
Both pre-moderator and moderator
were kept at a temperature of about \SI{6}{\kelvin}.

Every storage time setting was measured with 
three reactor pulses 
and then averaged.
To start the measurements, 
a TTL pulse approximately 
\SI{1.2}{\second} before
the start of the reactor pulse 
started the timing sequence.

\subsection{Filling optimization}

\begin{figure}[htb]
\begin{center}
\includegraphics[width=0.5\textwidth]{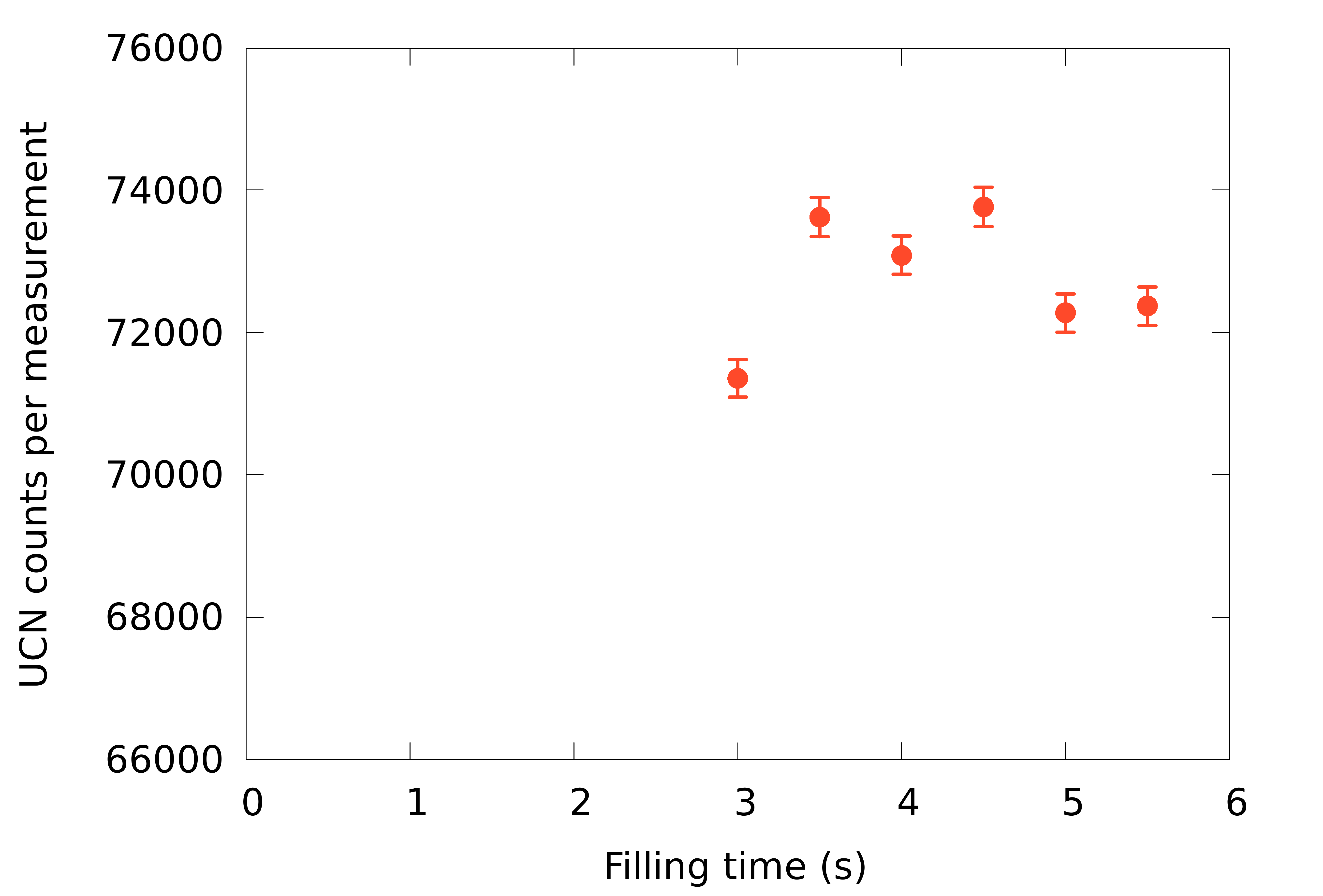}
\caption{
UCN counts per measurement 
after \SI{5}{\second} storage versus filling time
measured with vertical extraction.
}
\label{TRIGAFillingCurve111}
\end{center}
\end{figure}

\begin{figure}[htb]
\begin{center}
\includegraphics[width=0.5\textwidth]{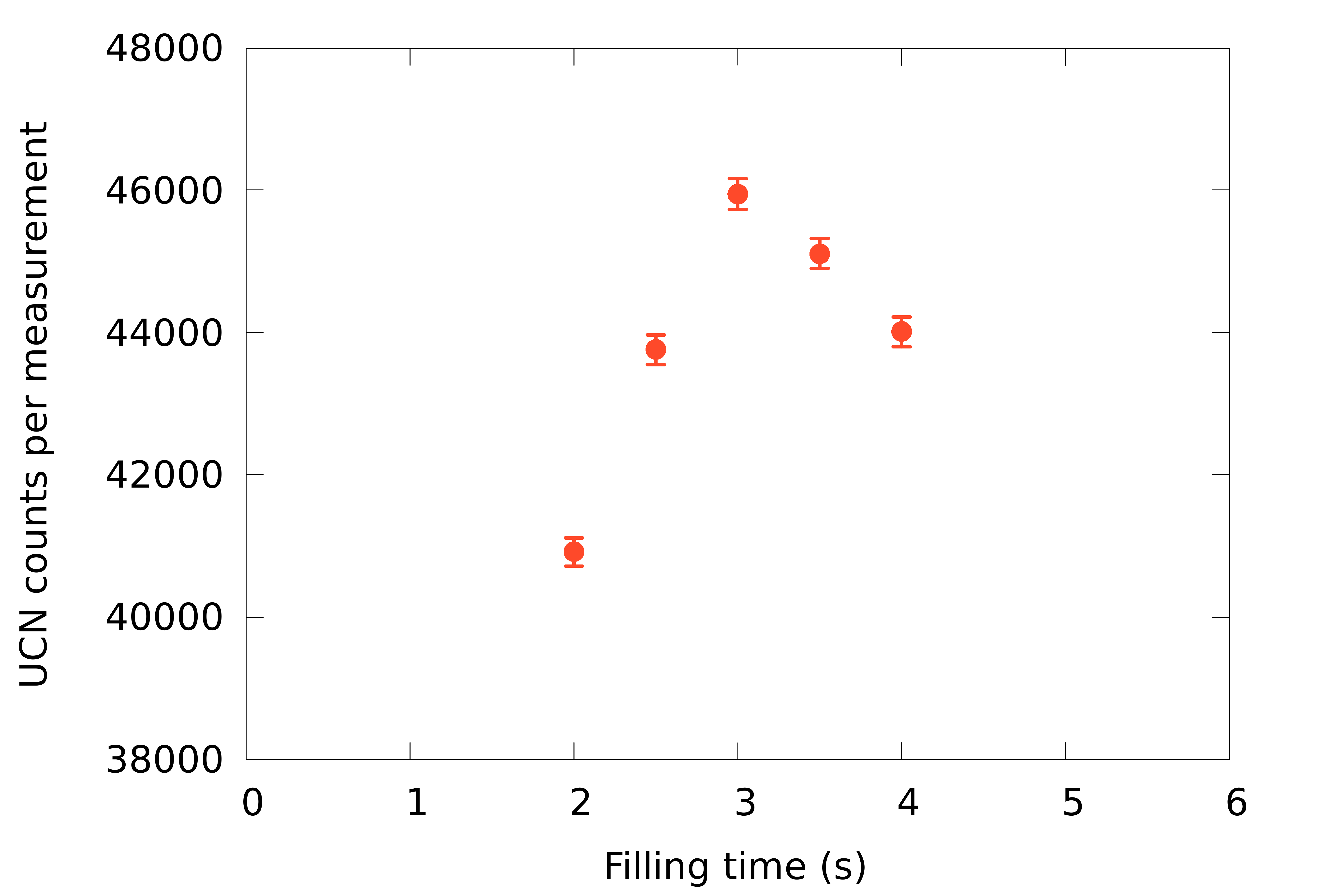}
\caption{
Same as Fig.~\ref{TRIGAFillingCurve111}
but measured with horizontal extraction.
}
\label{TRIGAFillingCurve2}
\end{center}
\end{figure}

The filling time was optimized as described in Sec.~\ref{StandardSequence}.
In contrast to all other sources 
the pulses at the TRIGA reactor are very short (\SI{30}{\milli\second} FWHM) 
resulting in short UCN pulses.
The reactor also delivers a pulse signal used to define t=0.
Results are shown for both extraction schemes
in Figs.~\ref{TRIGAFillingCurve111} 
and ~\ref{TRIGAFillingCurve2}. 
A filling time of \SI{4}{\second} was chosen for the vertical extraction
and \SI{3}{\second} for the horizontal extraction.

As the UCN 'pulse' for slow UCNs has broadened on its path from the production place
to the storage bottle, the maximum is less pronounced in the vertical extraction measurement,
but still much sharper than at other UCN sources.
The observed difference in optimal filling time can be qualitatively explained by the 
fact that the horizontal extraction has an additional threshold on detectable UCN energies
because of the Fermi potential of the detector entrance window (\SI{54}{\nano\electronvolt})~\cite{Golub1991},
and is therefore insensitive to very slow UCNs. 
Therefore, vertical extraction also shows a higher UCN yield.


\subsection{Storage measurements}

UCN storage measurements were performed 
with vertical and horizontal extraction at a height of 
\SI{130}{\centi\meter} above the beamport.
%

\begin{figure}[htb]
\begin{center}
\includegraphics[width=0.5\textwidth]{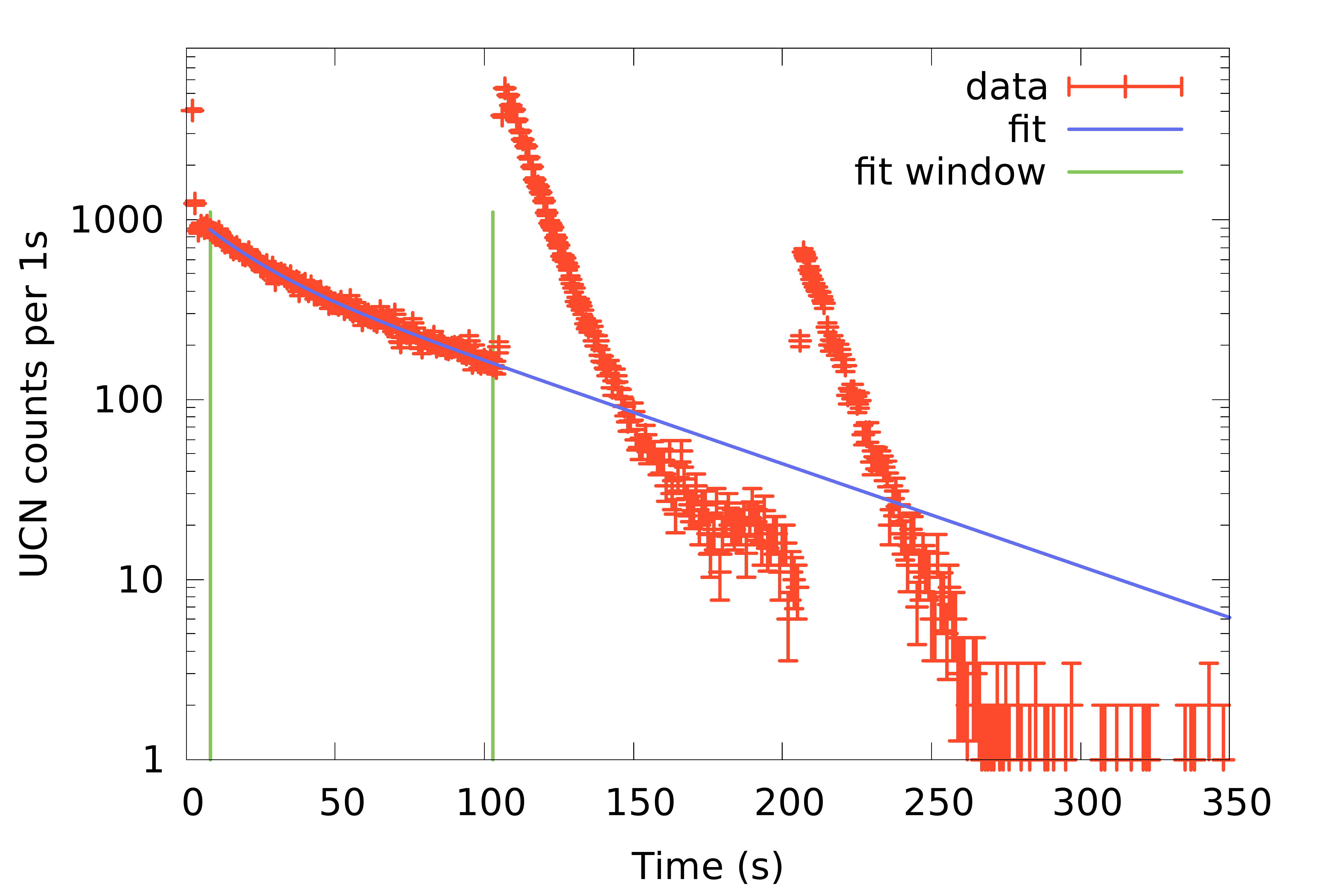}
\caption{
Leaking UCN counts versus time after the reactor pulse
measured with vertical extraction. 
Plotted is the sum of four measurements with \SI{100}{} and 
\SI{200}{\second} of storage time, therefore 2 
emptying peaks corresponding to these two times are visible. 
The fit window for the evaluation of the storage time constants
is indicated with vertical green lines.
}
\label{TRIGAFittedSpectrum}
\end{center}
\end{figure}

The storage time constants were extracted from fitting the leaking UCNs during
\SI{100}{} and \SI{200}{\second} storage measurements, as described 
in~Sec.~\ref{StandardAnalysis}. 
A time spectrum of UCN counts 
for storage times of \SI{100}{} and \SI{200}{\second}
is shown in Fig.~\ref{TRIGAFittedSpectrum}.
The fit results from the leakage are given 
in Tab.~\ref{TRIGAStorageTimeTable}.

\begin{table}[htb]
\begin{center}
\begin{tabular}{c|c|c|c|c|c}
	Extraction   &  $A_1$          &  $\tau_1$(s)         &  $A_2$        &  $\tau_2$(s)    & red.$\chi^2$            \\\hline
	horizontal   & \SI{246\pm40}{} &  \SI{12\pm4}{}  & \SI{443\pm40}{}  &  \SI{48\pm3}{}& \SI{0.86}{}   \\
	vertical     & \SI{336\pm70}{} &  \SI{18\pm4}{}  & \SI{545\pm80}{}  &  \SI{76\pm9}{}& \SI{0.92}{}   \\    
\end{tabular}
\caption{
Storage time constants and amplitudes 
from the fit to the leaking UCN time spectrum
(Fig.~\ref{TRIGAFittedSpectrum}).
}
\label{TRIGAStorageTimeTable}
\end{center}
\end{table}

\begin{figure}[htb]
\begin{center}
\includegraphics[width=0.5\textwidth]{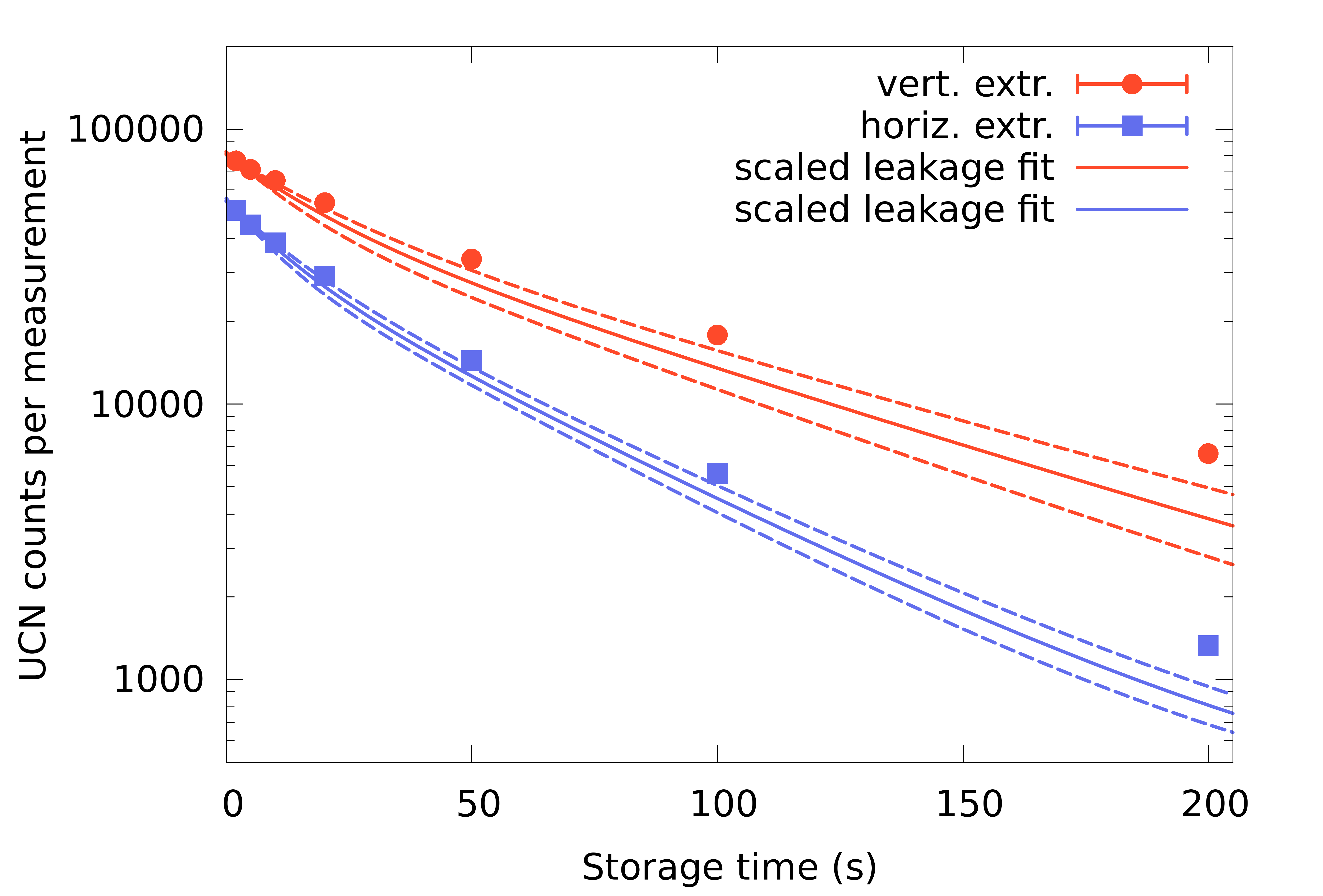}
\end{center}
\caption{
UCN counts per measurement after various storage times 
measured with filling times of 
\SI{4}{\second} (vertical extraction) and \SI{3}{\second} (horizontal extraction) 
at the TRIGA Mainz UCN source. 
The results of the fits to the leakage rates 
(see Tab.~\ref{TRIGAStorageTimeTable})
are indicated
by the continuous lines drawn with 1$\sigma$ error bands.
The data point at 200\,s was out of the fit window.
}
\label{TRIGAStorageCurves}
\end{figure}

The results of the storage measurements are shown in Fig.~\ref{TRIGAStorageCurves}
for horizontal and vertical extraction.
The lines indicate the results from the fit to the leaking UCNs
shown in Fig.~\ref{TRIGAFittedSpectrum} and
Tab.~\ref{TRIGAStorageTimeTable}.

\subsection{UCN density determination}

Table~\ref{TRIGAresultsTable} reports our results on 
the UCN counts observed in the \SI{2}{\second} measurements 
and the corresponding UCN densities.

In order to measure the amount of UCNs leaking
into the storage bottle during the counting time
for a measurement with a storage time of \SI{2}{\second},
the fast shutter at the beamport was closed 
up to the counting time and then opened.
%
%
%
UCNs still present in the beamtube of the source insert 
then leaked through the bottle shutter.
A total leakage 
as given in Tab.~\ref{TRIGAresultsTable}
was measured during the counting time
and subtracted from the data for the density measurements.

\begin{table}[htb]
\centerline{
\begin{tabular}{c|c|c|c|c}
 Extraction   & Net UCN             &    Subtracted        &  Density       \\
              & counts              &  leakage counts      &  (\SI{}{UCN\per\centi\meter^{3}})   \\ \hline
 horizontal   & \SI{51299\pm215}{}  &  \SI{722\pm30}{}     &   \SI{1.60\pm0.01}{} \\
 vertical     & \SI{77941\pm383}{}  &  \SI{1229\pm30}{}    &   \SI{2.43\pm0.02}{} \\
\end{tabular}}
\caption{
Results at the TRIGA Mainz:
Net UCN counts in \SI{2}{\second} storage measurements,
subtracted UCN leakage counts, 
and determined UCN density.		
The background due to leakage is very small.
}
\label{TRIGAresultsTable}
\end{table}

The 50\% higher UCN density measured with vertical extraction points again
at the large UCN population with energies below the 
threshold of the detector Al foil~\cite{Altarev2008}.

\section{Measurements at the solid D$_2$ source at the Paul Scherrer Institute}
\label{sec:PSI}

The UCN source at PSI is described in more detail in 
Refs.\cite{Anghel2009,Lauss2011,Lauss2012,Lauss2014,Goeltl2012,Becker2015}.
Operation permission was granted in 2011 and since then 
it has operated as a UCN user facility with 3 available
beamports. 
First UCN density measurements were reported in
\cite{Goeltl2012} with values of about
\SI{20}{UCN\per\centi\meter^{3}}.

The measurements described here
were performed at beamport West-1. 
Similar UCN intensities are provided at beamport
South, where the nEDM experiment is installed.

\subsection{Setup at PSI}

At the West-1 beamport, 
a \SI{1}{\meter} long glass guide with an 
inner diameter of \SI{180}{\milli\meter} 
and coated with \isotope{NiMo} 85/15~\cite{Blau2016}
was mounted, 
followed by a vacuum shutter of the same type as the 
beamline shutter. 
During the measurements at beam height, the storage bottle was 
directly connected to the second shutter.

\begin{figure}[htb]
\begin{center}
\includegraphics[width=0.5\textwidth]{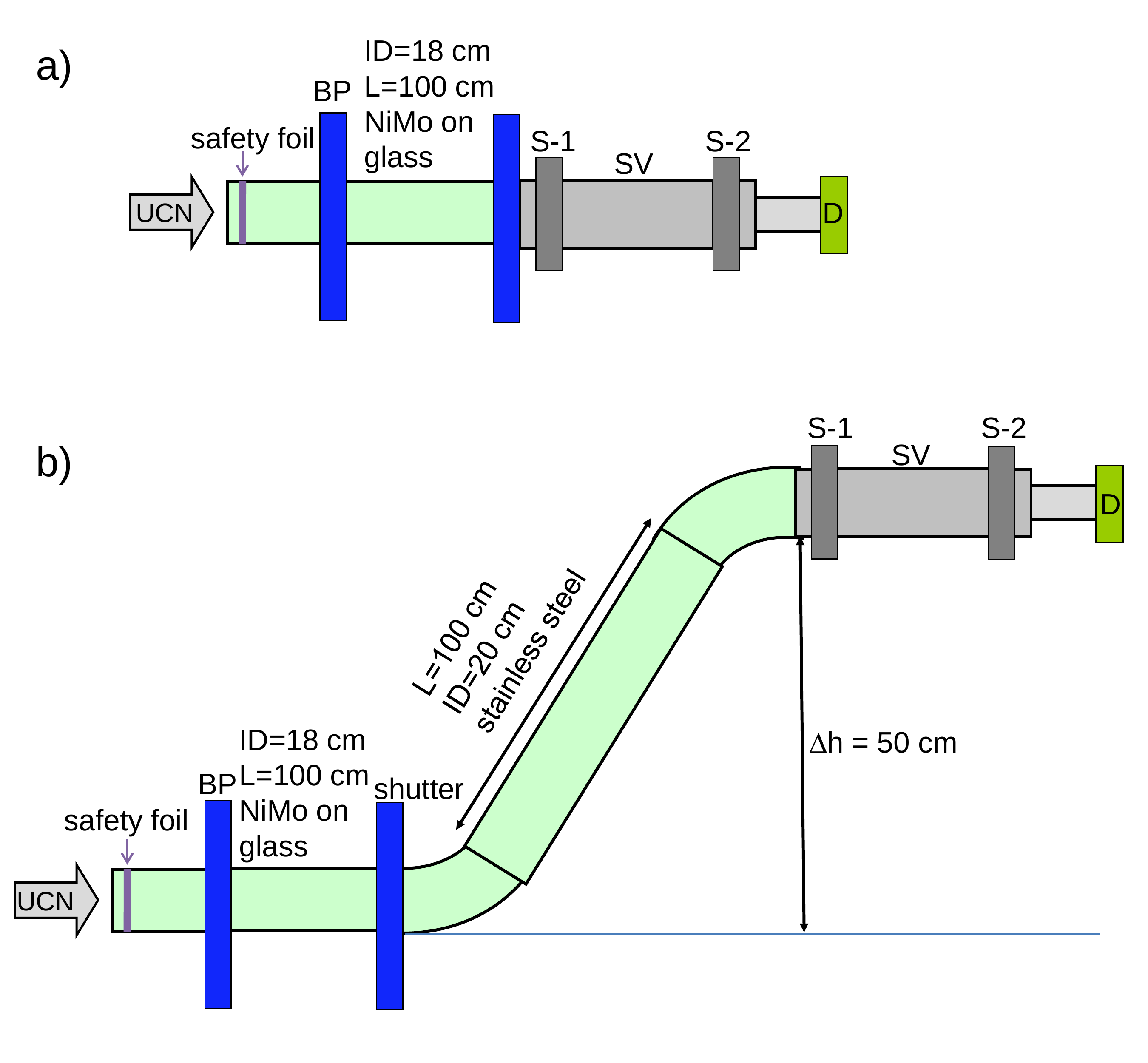}
\end{center}
\caption{
Sketch of the setup at PSI (not to scale): 
a) at the height of the beamport and 
b) at elevated height. 
Components as described in Fig.~\ref{SUN2setup}. 
In addition the location of the safety foil about 50\,cm before the beamport is indicated.
}
\label{PSIsetup}
\end{figure}

\begin{figure}[htb]
\begin{center}
\includegraphics[width=0.5\textwidth]{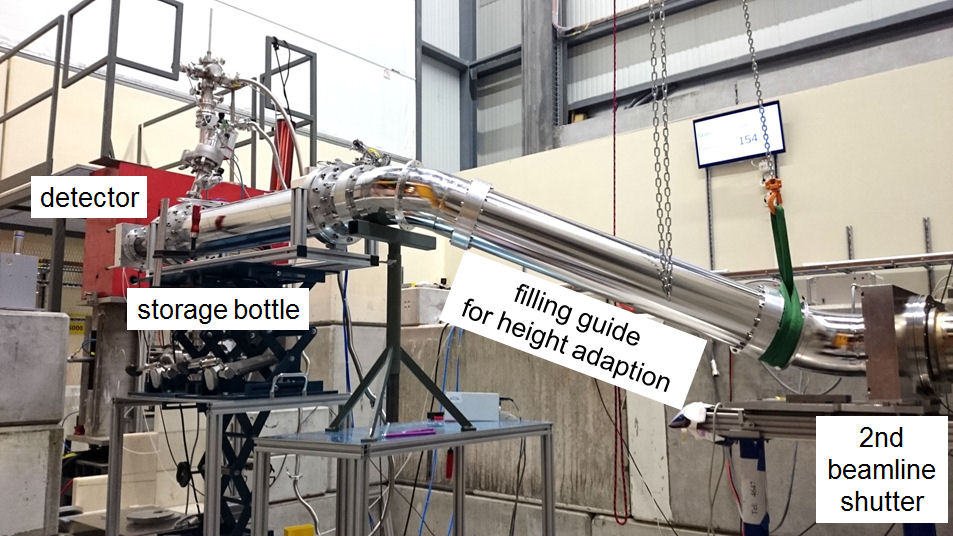}
\caption{PSI area West-1: Setup with additional stainless-steel guide section 
leading to the storage bottle mounted half a meter 
above the beam port with horizontal extraction to the detector.
}
\label{PSIWest1heightphoto}
\end{center}
\end{figure}

In front of 
the beamport on the UCN source side a 
100\,$\mu$m thick AlMg3 window serves as 
vacuum separation and safety foil, but
at the same time provides at low-energy UCN cut-off
due to its Fermi potential.
In order to compensate for this influence 
measurements were not only done at the height of the beamport 
but also \SI{500}{\milli\meter} higher.
An additional stainless-steel beamline section was attached,
made from two electro-polished \SI{45}{\degree} bends
(bending radius \SI{300}{\milli\meter}) 
and a tube of \SI{1}{\meter} length, 
all with inner diameter \SI{200}{\milli\meter}
and the same material as the storage bottle itself.
The storage bottle was then connected parallel to the beam port.
Both installations are sketched in Fig.~\ref{PSIsetup}.
The setup with the storage bottle and 
the additional guide section leading to
the higher position is shown in 
Fig.~\ref{PSIWest1heightphoto}.

\subsection{Operating conditions during the measurements}

The reported measurements at the elevated position used 
\SI{8}{\second} long proton beam pulses, 
which were repeated every
\SI{440}{\second} with a nominal proton beam current of 
\SI{2200}{\micro\ampere}, which is the standard
operating current at PSI's high-intensity proton accelerator.

Measurements at the height of the beamport 
were done using \SI{8}{\second} long
proton beam kicks every \SI{500}{\second} with a nominal proton beam current of 
\SI{2400}{\micro\ampere},
which is tested regularly on proton beam development days.
All described measurements were performed in December 2015.

The timing trigger for the measurements
was an accelerator signal arriving
\SI{1}{\second} before the rising edge of the proton
beam pulse.
The UCN source was operated with about 4.5\,kg of solid 
ortho-deuterium cooled to 5\,K 
after several conditioning cycles 
in Dec.~2015~\cite{Ries2016}.

\subsection{Determination of the optimal filling time}

\begin{figure}[htb]
\begin{center}
\includegraphics[width=0.5\textwidth]{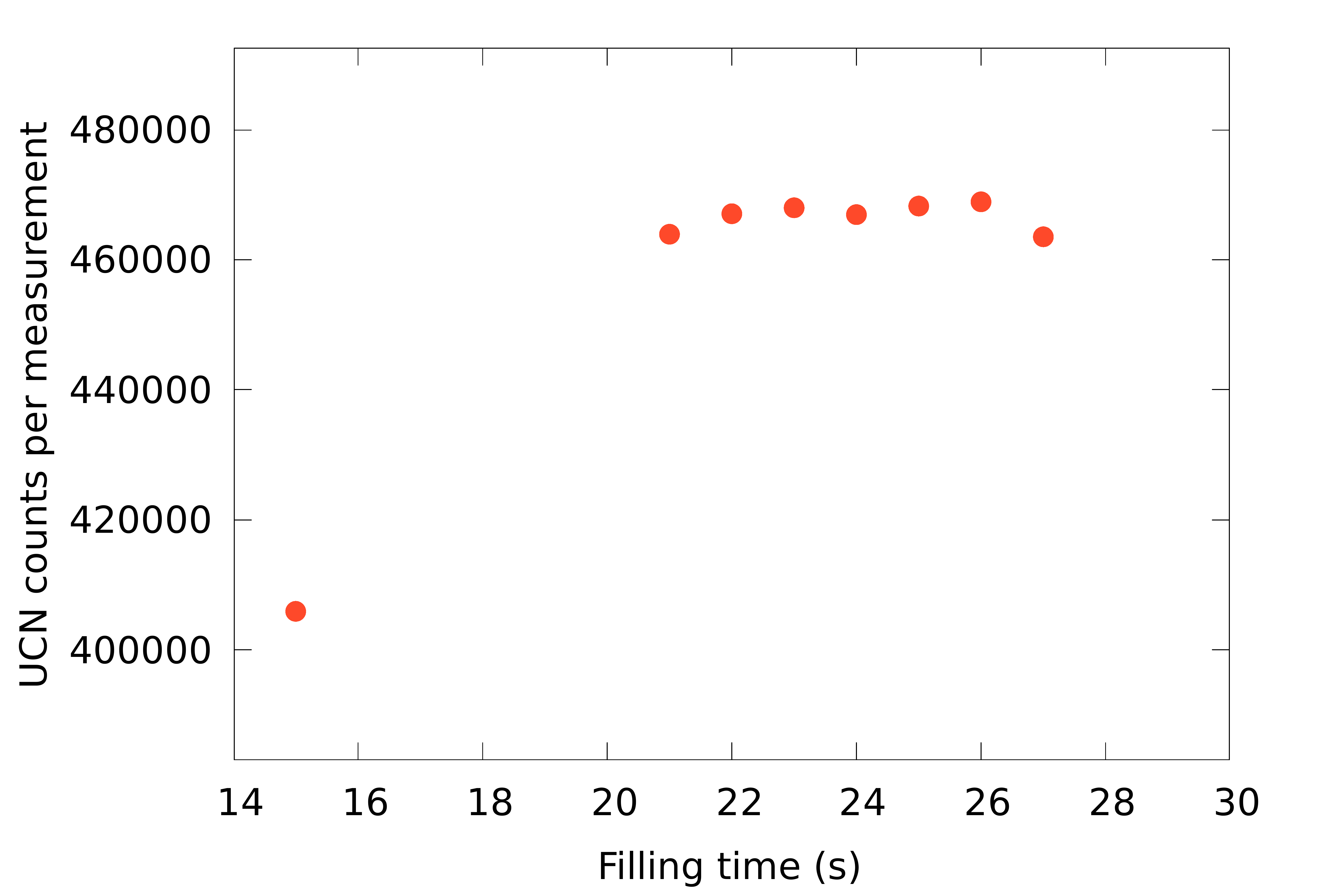}
\caption{
UCN counts per measurement 
after 5\,s of storage time 
for different filling times 
at beam height, at the PSI West-1 beamport,
measured with vertical extraction.
Statistical errors are below symbol size.
}
\label{PSIFillingCurve1}
\end{center}
\end{figure}

\begin{figure}[htb]
\begin{center}
\includegraphics[width=0.5\textwidth]{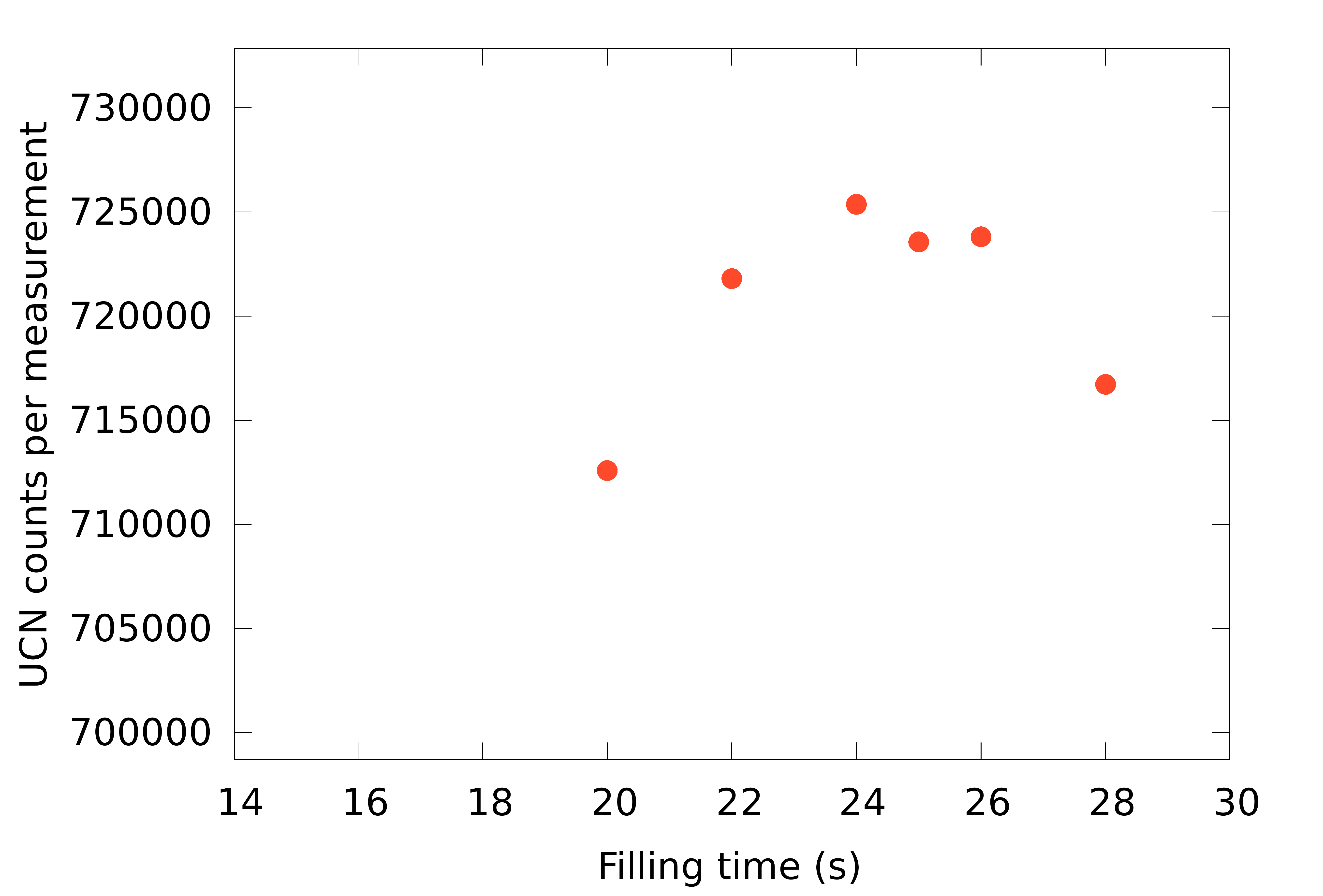}
\caption{
UCN counts per measurement 
after 5\,s of storage time for different filling times
at \SI{500}{\milli\meter} above beamport West-1, 
measured with vertical extraction.
Statistical errors are below symbol size.
}
\label{PSIFillingCurve2}
\end{center}
\end{figure}

The filling time was optimized as detailed in Sec.~\ref{StandardSequence}.
Time zero is defined via a proton beam signal which starts
the pulse sequence.
The resulting filling time plots are
shown in Figs.~\ref{PSIFillingCurve1} and~\ref{PSIFillingCurve2}.
At beam height, a filling time of \SI{26}{\second}
was chosen, while at the elevated position, 
a filling time of \SI{24}{\second} was chosen.

\subsection{Storage measurements}

\begin{figure}[htb]
\begin{center}
\includegraphics[width=0.5\textwidth]{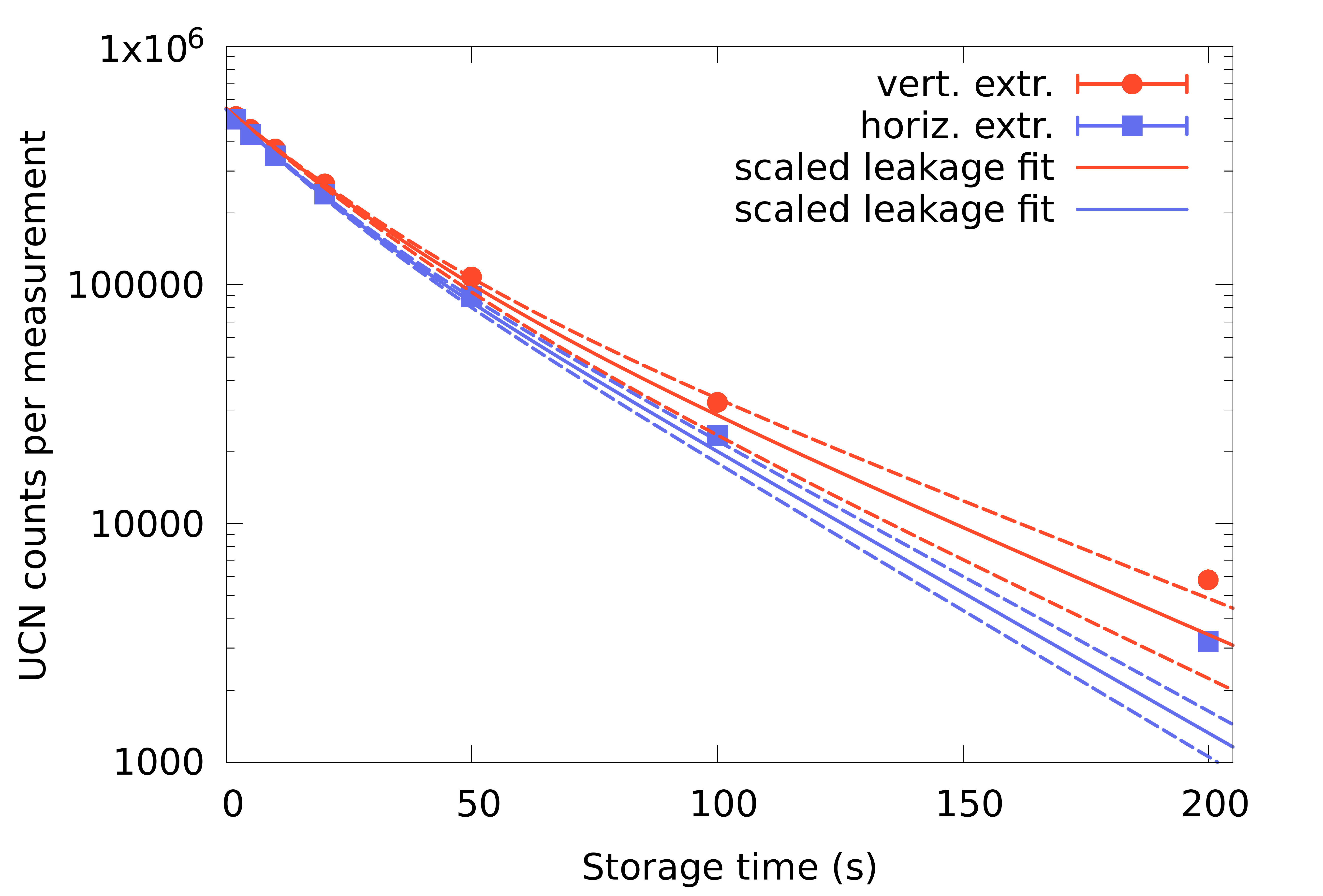} 
\caption{ 
UCN counts per measurement at beam height,
after various storage times, measured with filling times of 
\SI{26}{\second} (at beam height). 
The results of the fits to the leakage rates 
(see Tab.~\ref{PSIStorageTimeTable})
are indicated
by the continuous lines drawn with 1$\sigma$ error bands.
The data point at 200\,s was out of the fit window.
}
\label{PSIStorageCurve1}
\end{center}
\end{figure}

\begin{figure}[htb]
\begin{center}
\includegraphics[width=0.5\textwidth]{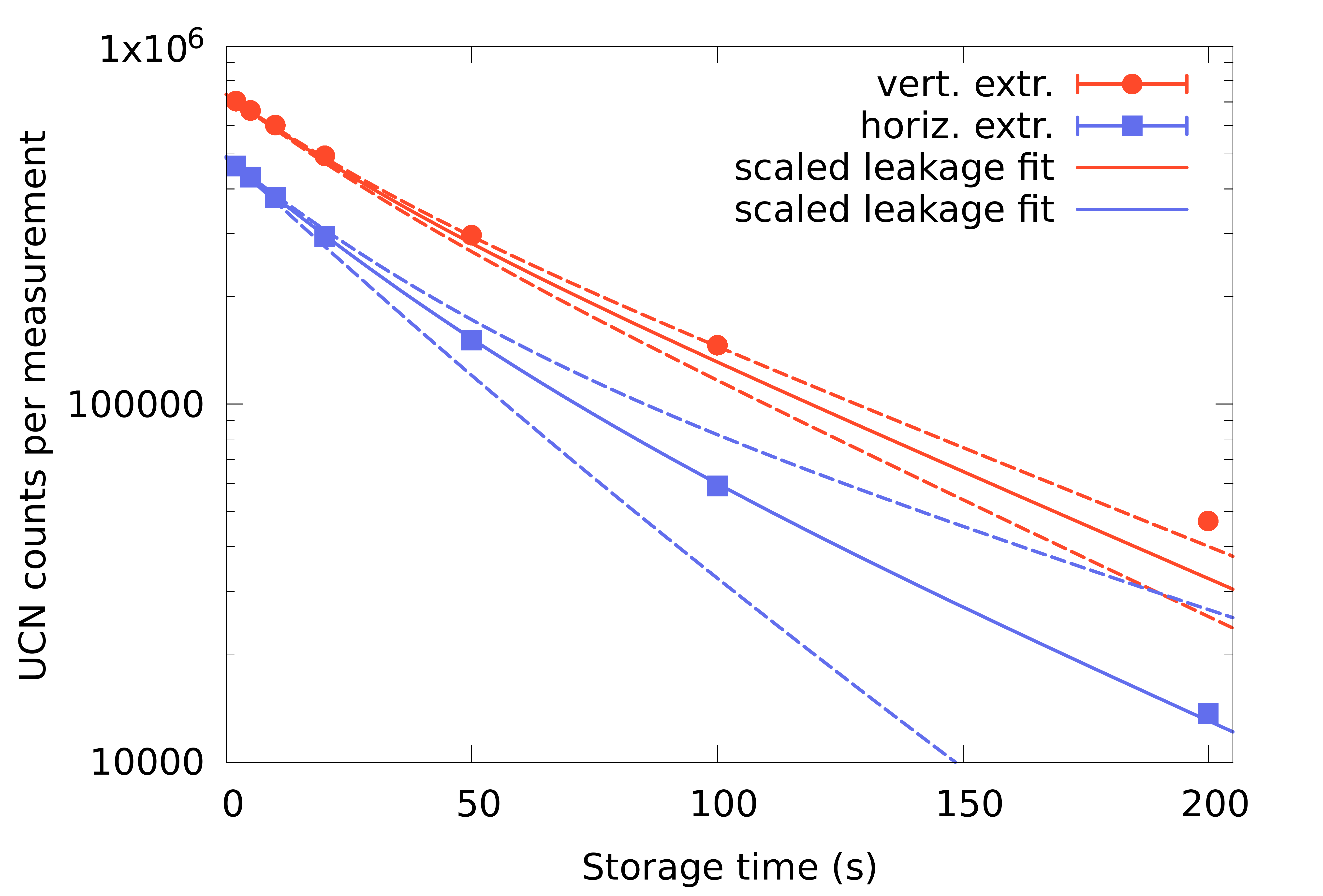}
\caption{Same as Fig.~\ref{PSIStorageCurve1}
but measured 
at \SI{500}{\milli\meter} height above beamport
with \SI{24}{\second} filling time.
The data point at 200\,s was out of the fit window.
The large error band in the horizontal extraction measurement is caused
by a strong correlation between the two storage time constants in the fit.
}
\label{PSIStorageCurve2}
\end{center}
\end{figure}

In order to keep leakage through shutter~1 to a minimum,
the beamline shutter was closed together synchronous to bottle shutter~1.
Storage measurements as described in Sec.~\ref{StandardSequence}
were conducted.
The resulting storage curves are shown in Figs.~\ref{PSIStorageCurve1}
and ~\ref{PSIStorageCurve2}.

\begin{figure}[htb]
\begin{center}
\includegraphics[width=0.5\textwidth]{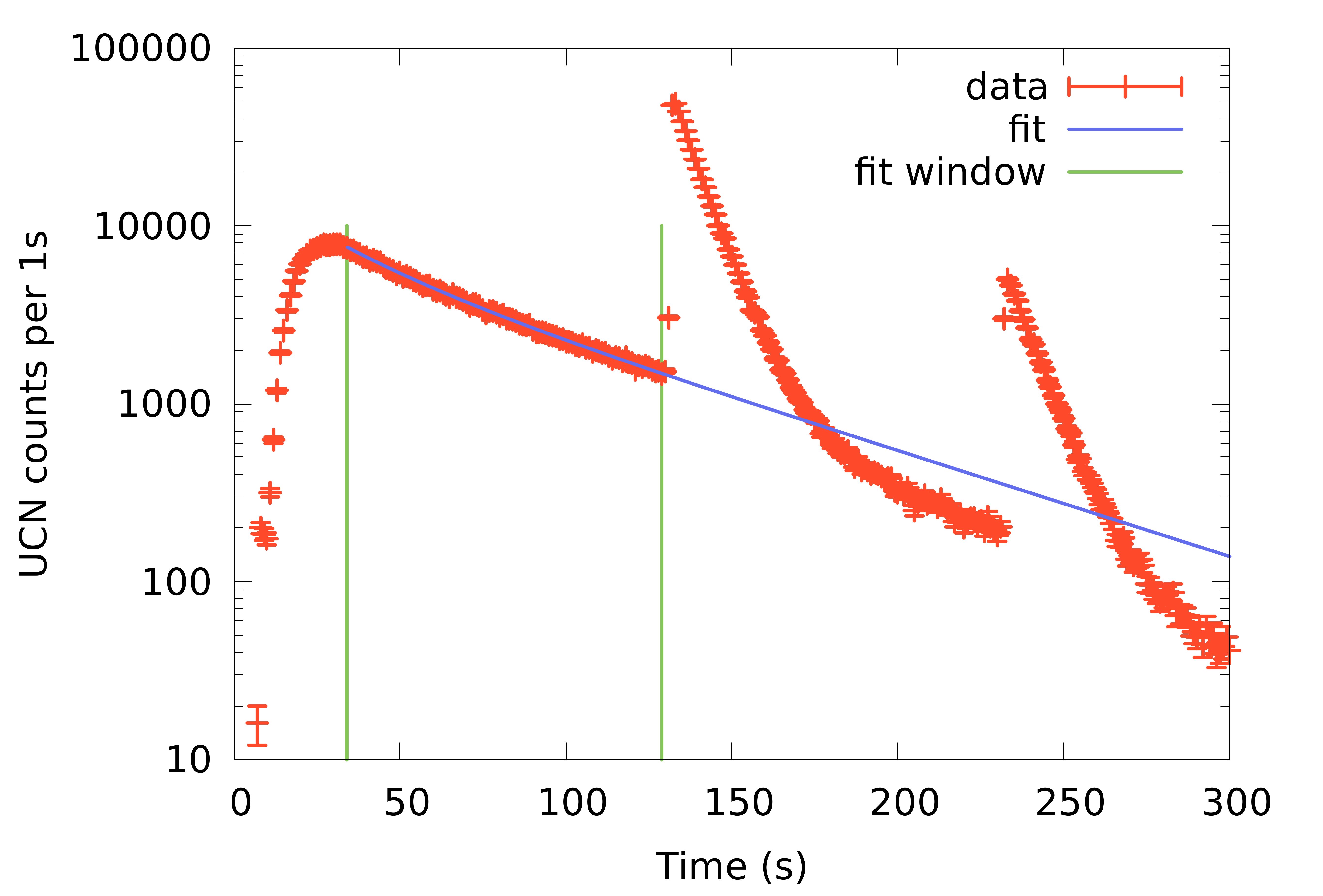}
\caption{
Time spectrum of UCN counts per 1\,s in the detector, 
measured at 
a height of \SI{500}{\milli\meter}, with vertical extraction.
Sum of four measurements
with \SI{100}{\second} and \SI{200}{\second} storage time,
therefore 2 emptying peaks are visible.
The double-exponential fit to the leakage rate of UCN during the storage time
(see Tab.~\ref{PSIStorageTimeTable})
is indicated (blue line), the vertical green lines show the fit window.
Time=0 is given by the accelerator signal before the main beam pulse.
}
\label{PSIFittedSpectrum}
\end{center}
\end{figure}

The storage time constants extracted from fitting the leakage rates during
\SI{100}{} and \SI{200}{\second} storage measurements, as described in 
\ref{StandardAnalysis}, are shown in Table.~\ref{PSIStorageTimeTable}.
An example for a time spectrum of UCN counts
together with the leakage rate fit is 
shown in Fig.~\ref{PSIFittedSpectrum}.

\begin{table}[htb]
\begin{center}
    \begin{tabular}{c|c|c|c|c|c|c}
   Height                & Extr. &  $A_1$  & $\tau_1$(s)   & $A_2$ & $\tau_2$ (s)  & red. \\
   (\SI{}{\milli\meter}) &       &         &         &       &        &  $\chi^2$       \\\hline
   0    & horiz.   & \SI{211\pm 26}{} &\SI{  14.5 \pm1.3}{} & \SI{275\pm 28}{} &  \SI{ 37.0\pm1.5}{} & \SI{1.00}{}\\
   0    & vert.    & \SI{394\pm 51}{} &\SI{  19.7 \pm1.5}{} & \SI{245\pm 53}{} &  \SI{ 49.0\pm4.6}{} & \SI{1.05}{}\\
   500  & horiz.   & \SI{153\pm 81}{} &\SI{  28.7 \pm7.6}{} & \SI{117\pm 82}{} &  \SI{ 72\pm26}{}    & \SI{1.13}{}  \\
   500  & vert.    & \SI{226\pm 66}{} &\SI{  23.7 \pm4.6}{} & \SI{520\pm 68}{} &  \SI{ 73.1\pm6.0}{} & \SI{1.11}{}\\
\end{tabular}
\caption{Parameters (amplitudes and storage times)
of the fit to the 
time distribution of UCN counts 
for the various measured positions at PSI.}
\label{PSIStorageTimeTable}
\end{center}
\end{table}


\subsection{UCN density determination}

Measured UCN counts, subtracted leakage, and densities are listed in 
Table~\ref{PSIresultsTable}.
Given errors are the standard deviations of
the mean averaged over typically three proton beam pulses per setting.
The given UCN densities correspond to the \SI{2}{\second} storage measurements.

An upper bound on the remaining UCN leakage 
from the UCN source to the detector during the counting time
was determined by performing a measurement
identically to a measurement with \SI{2}{\second} storage time, 
but with the shutter~1
manually set to be permanently closed. 
In addition, shutter~2 
was manually kept open during the filling period in order not to accumulate 
UCNs in the bottle. 
This measurement was then analyzed like a regular \SI{2}{\second} storage measurement.
%

\begin{table}[htb]
\centerline{
\begin{tabular}{c|c|c|c|c}
  Height         & Extr.   & Net UCN               &  Subtracted           & Density  \\
  (\SI{}{\milli\meter}) &  & counts                &  leakage counts       & (\SI{}{UCN\per\centi\meter^{3}}) \\\hline
      \SI{0}{}   & horiz.  & \SI{510687\pm1320}{}  &  \SI{15531\pm125}{}   & \SI{15.5\pm0.1}{} \\
      \SI{0}{}   & vert.   & \SI{523977\pm5198}{}  &  \SI{16625\pm129}{}   & \SI{15.8\pm0.2}{} \\
      \SI{500}{} & horiz.  & \SI{505138\pm711}{}   &   \SI{41345\pm203}{}  & \SI{14.5\pm0.1}{} \\
      \SI{500}{} & vert.   & \SI{767268\pm925}{}   &   \SI{66466\pm258}{}  & \SI{21.9\pm0.2}{} \\
\end{tabular}}
\caption{
PSI results: 
Net UCN counts in \SI{2}{\second} storage measurements,
subtracted UCN leakage counts, 
and determined UCN density.		
}
\label{PSIresultsTable}
\end{table}


The UCN density measured at the beamport shows almost no difference
between vertical and horizontal extraction.
Measurements 50\,cm higher show that the down-shift of 
the UCN energy spectrum helps to increase the UCN density
storable in a stainless-steel bottle by 50\%.


%

\vspace*{10mm}

\section{Comparison}

Due to the large variation in operating conditions 
a comparison between the sources is not straightforward.
We want to emphasize that the
UCN densities, as tabulated in the previous chapters,  
are calculated via
measured UCN counts divided by bottle volume
without extrapolations or efficiency corrections.
The given UCN densities are
fully independent of the given UCN storage time constants
for every source.

\begin{figure}[htb]
\begin{center}
\includegraphics[width=0.5\textwidth]{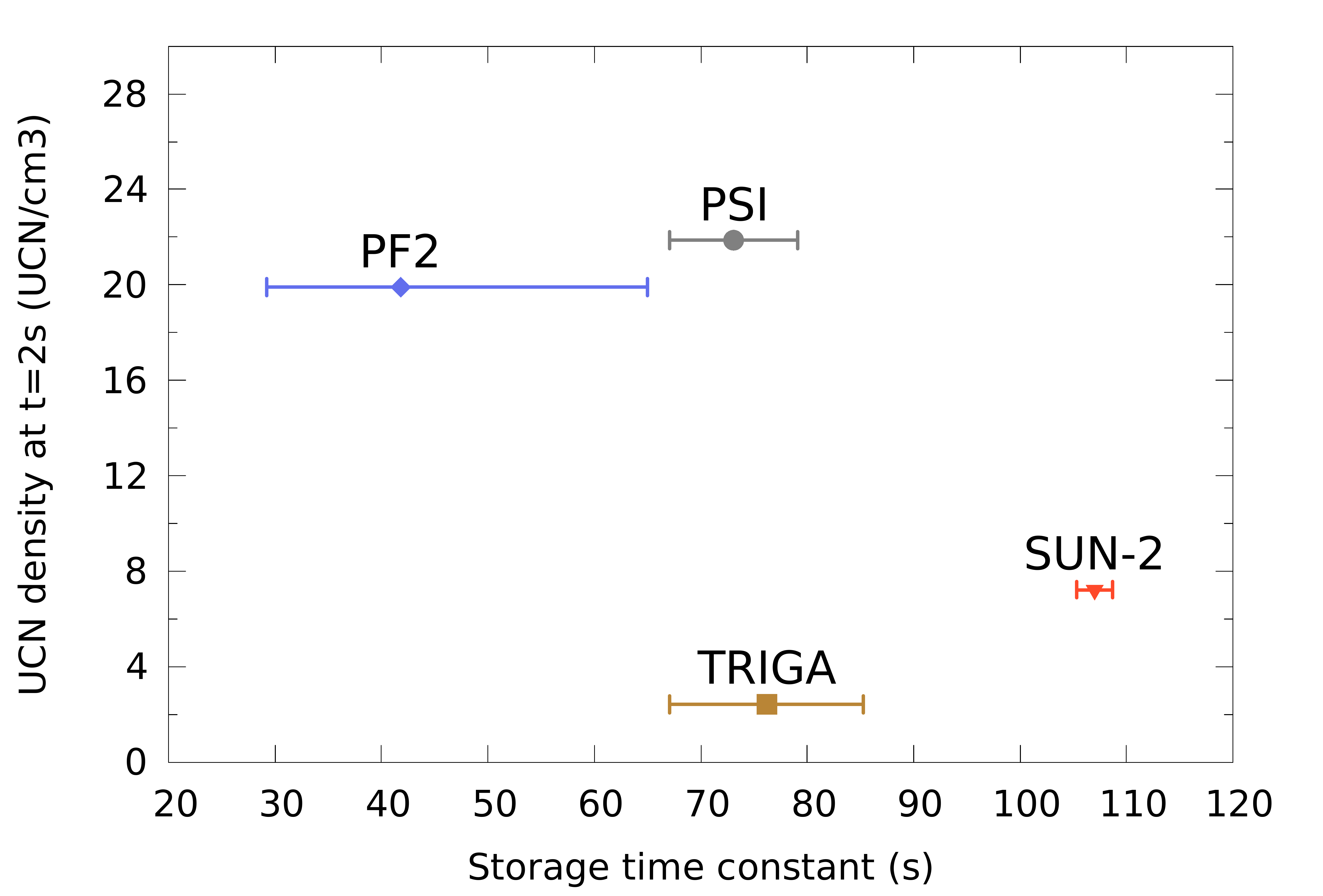}
\end{center}
\caption{
Measured largest UCN density in the standard storage bottle
at a given UCN source plotted versus the measured
storage time constant.
The measurement conditions are explained in the text.
The PF2 value is without safety foil, which is not
a standard user configuration.
Errors on the UCN density are smaller than symbol size.
}
\label{comparison-no-names}
\end{figure}

We have plotted the largest measured UCN density for a 
given UCN source 
together with the measured storage time constant in our
standard bottle in Fig.~\ref{comparison-no-names}.
Larger storage times are a clear indication
for lower UCN energies.
UCN energy spectra at the various
sources are largely different.

\begin{figure}[htb]
\begin{center}
\hspace*{-4mm}
\includegraphics[width=0.55\textwidth]{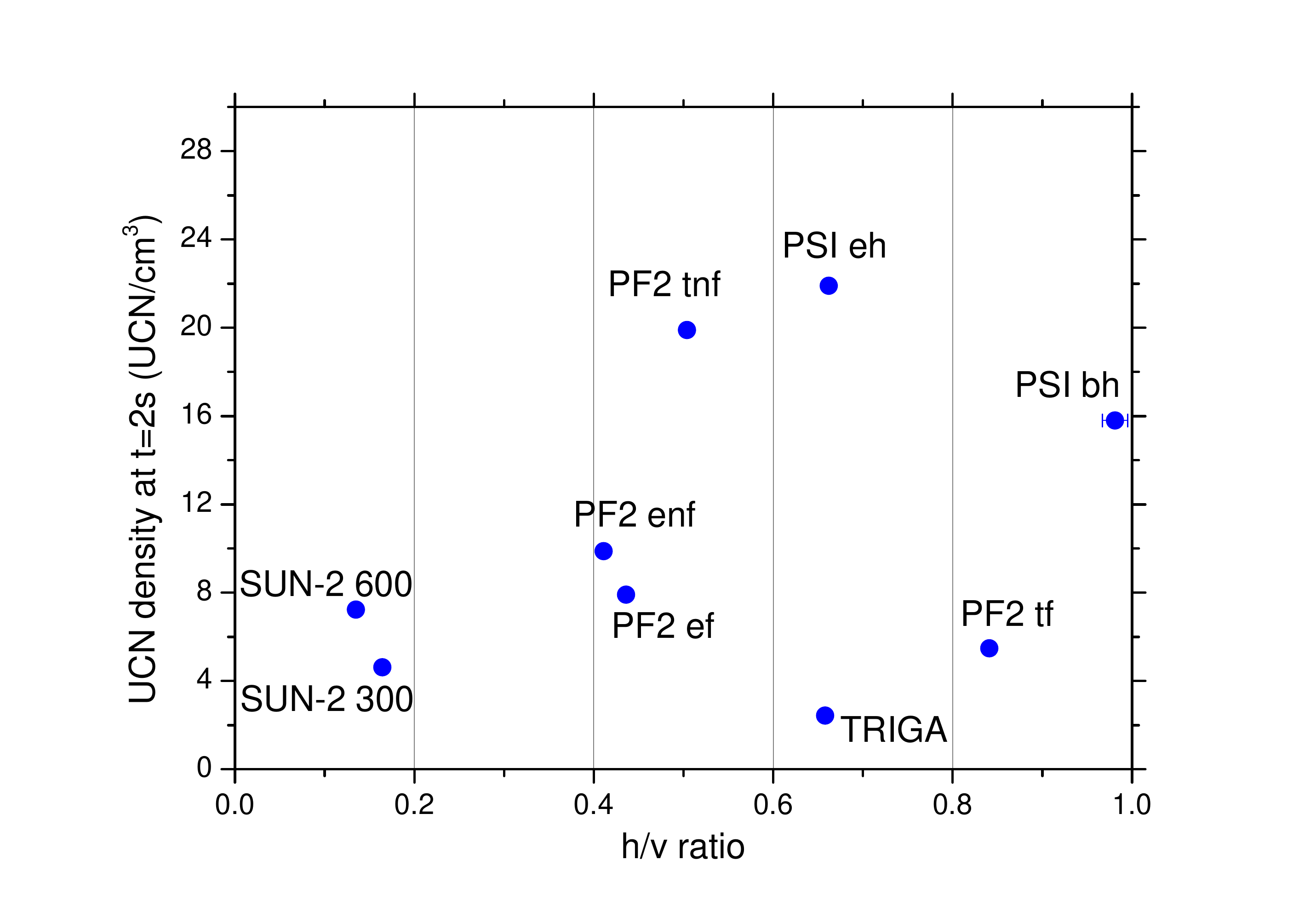}
\end{center}
\caption{
Calculated ratio of measured UCN density in horizontal
and vertical extraction, ``h/v ratio'', versus UCN density measured in 
vertical extraction for the given source in a 2\,s storage measurement.
Labels are explained in the text.
Errors are smaller than symbol size.
}
\label{hv-ratio-plot}
\end{figure}

Additional information on the UCN energy spectra
is obtained by the ratio between the  
UCN density measured with the detector in horizontal and vertical extraction position
(``h/v ratio''). 
The Cascade detector has an Al entrance foil
which acts in the horizontal extraction as energy barrier.
In vertical extraction UCN energies are increased by gravity.
Hence, the calculated ratio plotted in Fig.~\ref{hv-ratio-plot}
encodes the UCN fraction above and below the Al threshold energy.
The UCN sources operating with solid deuterium, 
PSI (Tab.~\ref{PSIresultsTable}), 
and TRIGA (Tab.~\ref{TRIGAresultsTable}),
show a comparable value of the h/v ratio.
One can see that for the PSI measurement
at beam height (PSI~bh)
the UCN are all above the 54\,neV threshold
of the Al safety foil in the beamline the UCN have to penetrate.
Hence, the ratio is consistent with 1.
Elevating the setup by 500\,mm (PSI~eh) decreases the UCN energies,
therefore the ratio is lower.
%
A similar behavior is visible at the PF2 source at turbine height 
with (PF2 tf) and without Al foil (PF2 tnf)
(from Tab.~\ref{PF2resultsTableDown}).
Moving up on the EDM platform at PF2 also shifts the UCN energy spectrum 
downwards, which is indicated in the lower h/v ratio 
in comparison to the 
measurement at turbine height
for the measurements with Al safety foil (PF2 ef) 
and without safety foil (PF2 enf)
(from Tab.~\ref{PF2resultsTablePlat}).
%
Measurements at the SUN-2 source 
are indicated with accumulation time of 300\,s (SUN-2 300)
and 600\,s (SUN-2 600) from Tab.~\ref{SUN2resultsTable}.
The very low h/v value indicates that the UCN spectrum
has by far the lowest mean energies of the measured sources
with a large fraction of UCNs below 54\,neV.
At SUN-2 one has to keep in mind the fast deterioration of the 
source performance due to storage bottle outgassing.
The measurements with 300\,s accumulation time were all performed later
than the one with 600\,s, hence the performance was already deteriorated.

\section{Summary}

In this study we compared all ultracold neutron sources
operating at different institutions.
We have established one standard method for comparison 
of UCN densities, which emphasizes storage chambers of sizes
typically used in room-temperature neutron EDM searches.

Major efforts have been undertaken in the last decade 
and continue today 
to develop improved UCN sources.
The UCN density performance of all UCN sources 
is so far comparable and within a factor of 10.
The observed h/v ratios and storage time constants at the different UCN sources are consistent 
with similar UCN energy spectra delivered by the solid-deuterium sources, 
and a lower-energy spectrum from the superfluid-helium source as expected 
due to the Fomblin coating of the converter vessel.
All the considered UCN sources have their respective merits
and are very useful for various purposes and goals.

This work has strengthened the sound basis for
co-operation between UCN sources located
at institutions in different countries
in order to
facilitate joint progress in the field.
When in the future new UCN sources will become operating,
the here reported standard could be used to allow for comparison
with the present work.

\vspace*{4mm}

\begin{acknowledgments}
This work is part of the PhD thesis of Dieter Ries.
D.R. acknowledges the support and hospitality
during the measurements at ILL, Mainz University and
Los Alamos National Lab.
ILL acknowledges support by the French Agence Nationale de la Recherche (ANR) 
for R\&D work based on the SUN-2 prototype towards the user facility SuperSUN.
Johannes Gutenberg University of Mainz (JGU) acknowledges 
the Stiftung Rheinland-Pfalz f\"ur Innovation (962-386261/993),
the Cluster of Excellence PRISMA (DFG 1098), and 
the mechanical and electronic workshops at the 
Institutes of Nuclear Chemistry and of Physics
of JGU.
ETH acknowledges the support by the Swiss National Science Foundation
Projects 200020\_149211 and 200020\_162574. 
PSI acknowledges the support by the Swiss National Science Foundation
Projects 200020\_137664 and 200020\_149813, 200020\_163413. 
We acknowledge 
the proton accelerator operations section, 
and all people who have been contributing to the UCN source construction 
and operation at PSI, 
and especially the BSQ group which has been operating the PSI source, namely
B.~Blau, A.~Anghel and P.~Erisman. 
There would be no storage vessel without the support of F.~Burri and M.~Meier.
\end{acknowledgments}

%


\end{document}